\documentclass[pdftex,pra,aps,twocolumn,twoside]{revtex4-2}
\usepackage[utf8]{inputenc}
\usepackage{amsmath, amsthm}
\usepackage{float}
\makeatletter
\let\newfloat\newfloat@ltx
\makeatother
\usepackage{algorithm}
\usepackage{algpseudocode}
\usepackage{tikz}
\usepackage[hidelinks]{hyperref}
\usepackage{braket}
\usepackage{float}
\usepackage{color, colortbl}
\usepackage{breqn}
\usepackage{caption}
\usepackage{subcaption}
\usepackage{cleveref}
\usepackage{autonum}
\usepackage{flexisym}
\usepackage{array}
\usepackage{ragged2e}

\newenvironment{mytabular}[1][1]{%
  \begin{tabular}%
}{%
  \end{tabular}
}
\theoremstyle{plain}
\newtheorem{theorem}{Theorem}

\newtheorem{definition}[theorem]{Definition}

\newtheoremstyle{note}{\topsep}{\topsep}{\slshape}{}{\scshape}{}{ }{}
\theoremstyle{note}

\algrenewcommand\algorithmicrequire{\textbf{Input:}}
\algrenewcommand\algorithmicensure{\textbf{Output:}}

\definecolor{Gray}{gray}{0.85}
\newcolumntype{a}{>{\columncolor{Gray}\centering\arraybackslash}m}
\newcolumntype{k}{>{\centering\arraybackslash}m}
\newcommand\+{{\footnotesize \raisebox{0.3ex}{+}}}
\begingroup\makeatletter
\catcode`,=\active
\global\let\breqn@comma,
\protected\gdef,{\ifmmode\expandafter\breqn@comma\else\expandafter\active@comma\fi}
\endgroup
\begin{document}
\title{Ultrafast Hybrid Fermion-to-Qubit mapping}
\author{Oliver O'Brien, Sergii Strelchuk}
	\affiliation{DAMTP, Centre for Mathematical Sciences, University of Cambridge, Cambridge CB30WA, UK}
\begin{abstract}

Fermion-to-qubit mappings play a crucial role in representing fermionic interactions on a quantum computer. Efficient mappings translate fermionic modes of a system to qubit interactions with a high degree of locality while using few auxiliary resources.  We present a family of locality-preserving fermion-to-qubit mappings that require fewer auxiliary qubits than all existing schemes known to date. One instance requires only 1.016 qubits-per-fermion compared to 1.25 for the best-known locality-preserving mapping by Y.-A. Chen and Y. Xu [PRX Quantum 4, 010326 (2023)]. Our family of mappings (parameterised by integer $n$) establishes a direct trade-off between the number of auxiliary qubits ($\frac{1}{n^2}$) and the circuit length ($O(\log n)$). Furthermore, we present a non-local variant that combines the strengths of the Jordan-Wigner and Bravyi-Kitaev mappings to give 98\% shorter circuits than the Jordan-Wigner mapping. This is achieved by applying seemly incompatible mappings at different scales, making it possible for their respective strengths to complement each other.

\vspace{1.12cm}
\end{abstract}
\maketitle

\section{Introduction}
Simulating quantum systems is one of the most promising applications of quantum computing in the near future~\cite{feynman2018simulating,lloyd1996universal,ortiz2002simulating}. The class of fermionic systems is particularly important and simulating them has an impressive range of applications: from problems in quantum chemistry~\cite{mcardle2020quantum,reiher2017elucidating}, to the Hubbard model in condensed matter physics~\cite{abrams1997simulation}, to Lattice Gauge Theories \cite{banuls2020simulating,chakraborty2022classically} and beyond. Classical methods applied to these problems do not scale favourably with the system size, whereas quantum techniques (e.g. VQE~\cite{mcclean2016theory}, Trotterization~\cite{tranter2019ordering} and others \cite{bravyi2002fermionic,friesner2005ab,ortiz2001quantum}) open up efficient avenues for studying static and dynamic properties of fermionic systems. In recent years, there has been a considerable experimental effort to simulate small fermionic systems on a quantum computer~\cite{barends2015digital,kokail2019self}.

 One of the crucial challenges facing these simulations is encoding the anti-commuting fermionic variables as operators acting on the qubit degrees of freedom. A desirable transformation would map interactions in the fermion picture to qubits resulting in (quasi-) local operators for the latter side. Furthermore, it should be engineered in a way to substantially reduce gate overhead in a circuit model -- leading to faster simulation times and more robustness with respect to errors.

There is currently no known fermion-to-qubit mapping that preserves perfect locality and consumes no auxiliary resources (in the form of qubits or the extra gate overhead) for an arbitrary interaction graph. However, in the last decade there has been remarkable progress towards designing efficient mappings~\cite{setia2019superfast,havlivcek2017operator,bravyi2002fermionic,seeley2012bravyi, derby2021compact,chen2022equivalence, steudtner2019quantum} and exploring new ideas that would improve the existing ones~\cite{chiew2021optimal}. A summary of select mappings (including the most efficient ones) can be found in Table~\ref{tab:performance}. These mappings fall into two main categories: non-local mappings which require no additional resources but produce gate counts dependant on the total number of fermionic modes; local mappings which use additional resources (ancilla qubits) to produce constant gate counts independent of the number of fermionic modes. The best-known local qubit mapping is due to Chen and Xu~\cite{chen2022equivalence}. It uses the fewest qubits per fermion mode (1.25) while producing Hamiltonians with $k$-local average and maximum Pauli string, for constant $k$. 

In this work, we introduce a family of parametrized Hybrid mappings (Section~\ref{section:Hybridmapping}) that combine the relaxed connectivity constraints of the Jordan-Wigner mapping and the increased locality of the Bravyi-Kitaev mapping to produce drastically reduced gate counts~\cite{tranter2018comparison}. In the context of limited qubit connectivity (Section~\ref{section:connectivity}) the Hybrid mapping produces gate counts that scale with $\frac{N}{2n}$ compared to $\frac{N}{2}$ for the Jordan-Wigner and Bravyi-Kitaev mappings on an $N\times N$ lattice (where $n \ll N$). In the regime of all-to-all connectivity the Hybrid mapping gate count scales with $\frac{N}{2n^2}$ (Section~\ref{section:hybridanalysis}) even further outperforming the unaffected Jordan-Wigner mapping. However, in the absence of connectivity constraints the Bravyi-Kitaev mapping gate count scales with $O(\log N)$. Despite this our numerical analysis in Section~\ref{section:numerical} shows the Hybrid mapping still results in lower gate counts than the Bravyi-Kitaev mapping for small lattices (e.g. lattices smaller than $164 \times 164$ for $n=4$).

We also construct a Hybrid\+ family of parametrized local mappings (Section~\ref{section:auxHybridmapping}) which only require $1+\frac{1}{n^2}$ qubits per fermionic mode. Our construction interpolates between the Auxillary Qubit mapping~\cite{steudtner2019quantum} and the Bravyi-Kitaev mapping by setting $n=1$ for the former and $n=N$ for the latter. For all the intermediate $n$, our construction results in novel highly efficient mappings that produce a dramatic reduction in the number of auxiliary qubits required, while simultaneously generating local Pauli strings. We illustrate examples for $n=4$ and $n=8$ which demonstrate 1.0625 and 1.0156 qubit-to-fermion ratios compared to the best existing local mapping \cite{chen2022equivalence} which achieves a 1.25 qubit-to-fermion ratio.

The key insight of our work is that it is possible to apply seemly incompatible mappings on different scales such that their relative strengths complement each other. We combine the efficiency of the Bravyi-Kitaev mapping with the low connectivity restraints of the Jordan-Wigner mapping to produce a far superior Hybrid mapping. On a small scale (within each cell) we use the Bravyi-Kitaev mapping (in few fermion regime when it's connectivity restraints are not too large), but when we zoom out to a larger scale we use the Jordan-Wigner mapping between the cells. By varying the scale at which each mapping is dominant we are able to find an optimal mapping for a given quantum computer. 

\begin{table}
\centering
\small
    \begin{tabular}{k{5.5em} k{4.6em} k{4.5em} k{4.6em} k{6.5em}}
        \hline \hline
         & Qubit to fermion ratio  &Max Pauli string & Avg. Pauli string & Avg. interaction qubit count on square-lattice architecture\\ \hline \rowcolor{Gray}
        
        \begin{mytabular}[1]{c}Jordan-\\Wigner\cite{jordan1993paulische}\end{mytabular}&1&$2N$&$\frac{N}{2} +\frac{3}{2}$&$\frac{N}{2} +\frac{3}{2}$ \\
        \begin{mytabular}[1]{c}Bravyi-\\Kitaev\cite{bravyi2002fermionic}\end{mytabular}  &
        1&$\mathcal{BK}_{N^2}$&$\overline{\mathcal{BK}}_{N^2}$&$O(\frac{N}{2})$*\\\rowcolor{Gray}
        \begin{mytabular}[1]{c}$n\times n$\\{\bf  Hybrid}\end{mytabular}&$1$&\begin{tabular}{c}$\frac{2N}{n} +$\\ $2\mathcal{BK}_{n^2}$\end{tabular}&\begin{tabular}{c}$\frac{N}{2n^2} +$\\ $\frac{n+1}{n}\overline{\mathcal{BK}}_{n^2}$\end{tabular}&$O(\frac{N}{2n}+\frac{n}{2})$\\
        \begin{mytabular}[1]{c}Derby-\\Klassen\cite{derby2021compact} \end{mytabular}&1.5&$3$&$3$&$3$\\\rowcolor{Gray}
        \begin{mytabular}[1]{c}Super-\\compact\cite{chen2022equivalence} \end{mytabular}&1.25&$6$&$O(1)$&$O(1)$\\
        AQM\cite{steudtner2019quantum}&$2-\frac{1}{N}$&$6$&$O(1)$&$O(1)$\\\rowcolor{Gray}
        \begin{mytabular}[1]{c}$n\times n$\\{\bf  Hybrid\+}\end{mytabular}&$1+\frac{1}{n^2}$&\begin{mytabular}[1.2]{c}$3 +$\\ $2\mathcal{BK}_{n^2}$\end{mytabular}&\begin{tabular}{c}$\frac{2}{n} +\frac{1}{n^2}+$\\ $\frac{n+1}{n}\overline{\mathcal{BK}}_{n^2}$\end{tabular}&$O(\frac{n}{2})$*\\ \hline
    \end{tabular}
    \caption{\label{tab:performance} { Comparison of the performance (on an $N \times N$ square-lattice fermionic interaction graph) of select families of fermion-to-qubit mappings including state-of-the-art alongside our novel mapping $n\times n$ Hybrid and Hybrid\+ mappings for $n \ll N$.  $\mathcal{BK}_m$ and $\overline{\mathcal{BK}}_m$ represent the $O(\log m)$ maximum and average Pauli-weight of fermionic operators under the Bravyi-Kitaev mapping on $m$ fermionic modes. For each hybrid mapping, $n$ is fixed. For $n=1$ the Hybrid mapping is equivalent to the Jordan-Wigner mapping, and the Hybrid\+ mapping is equivalent to AQM. For $n=N$ both are equivalent to the Bravyi-Kitaev mapping. Intermediate values of $n$ combine the strengths of these mappings to give improved performance. The optimal choice of $n$ is dependant upon $N$ and must be found numerically (Figures~\ref{fig:vary_cell_sizes_32},\ref{fig:vary_cell_sizes_64}). For $N=64$, the optimal choice is $n=8$. The $8 \times 8$ Hybrid mapping gives 98\% shorter Pauli-strings than the Jordan-Wigner mapping, and the $8 \times 8$ Hybrid\+ mapping only requires 1.016 qubits per fermion. This is 94\% fewer ancillae than the super-compact mapping (which requires the fewest ancillae of all known local mappings). *These values are estimated using numerical evaluation.} } 
\end{table}
\section{Preliminaries}
 In this section, we introduce the mappings our Hybrid schemes extend: the Jordan-Wigner Mapping (Section~\ref{section:jordan-wigner}), the Bravyi-Kitaev Mapping (Section~\ref{section:bravyi-kitaev}), and the Auxillary Qubit Mapping (Section~\ref{section:auxillaryMappings}). We also outline the Steiner Tree Solver (Section~\ref{section:SCIPJACK}) used to analyse the performance of our mappings in the presence of limited qubit connectivity.
\subsection{Jordan-Wigner Mapping}
\label{section:jordan-wigner}
The Jordan-Wigner transform \cite{jordan1993paulische} is arguably the oldest known fermion-to-qubit mapping. It stores the occupation number of the $i$-th fermionic mode in the $i$-th qubit. Therefore, all the non-local behaviour (the parity information) has to be encoded in the operators \cite{chiew2021optimal}:
\begin{align}
        a_i \rightarrow \left( \bigotimes_{k=1}^{i-1} Z_k \right) \sigma_i^-\\
        a^{\dagger}_i \rightarrow \left( \bigotimes_{k=1}^{i-1} Z_k \right) \sigma_i^+
\end{align}
where
\begin{align}
        \sigma^-_i = \frac{1}{2} (X_i + i Y_i) = \ket{0}_i \bra{1}_i\\
        \sigma^+_i = \frac{1}{2} (X_i - i Y_i) = \ket{1}_i \bra{0}_i 
\end{align}
The $\bigotimes_{k=1}^{i-1} Z_k$ term is necessary to apply the parity phase shift $(-1)^{\sum_{k=0}^{i-1} f_k}$, where $f_k$ are defined below. These operators act in exactly the same way on a qubit spin basis as the fermionic creation and annihilation operators act upon the Fock basis in Eqns~\eqref{secondQuant1} and~\eqref{secondQuant2}.

\begin{multline}
    \label{secondQuant1}
    a_p \ket{f_0,\cdots, f_i, \cdots, f_{M-1}} =\\ \delta_{f_p, 1} (-1)^{\sum_{i=0}^{p-1} f_i} \ket{f_0,\cdots, f_p \oplus 1, \cdots, f_{M-1}}
    \end{multline}
\begin{multline}
    \label{secondQuant2}
    a_p^{\dagger} \ket{f_0,\cdots, f_i, \cdots, f_{M-1}} =\\ \delta_{f_p, 0} (-1)^{\sum_{i=0}^{p-1} f_i}
    \ket{f_0,\cdots, f_p \oplus 1, \cdots, f_{M-1}}
    \end{multline}
\subsection{Bravyi-Kitaev Mapping}
\label{section:bravyi-kitaev}
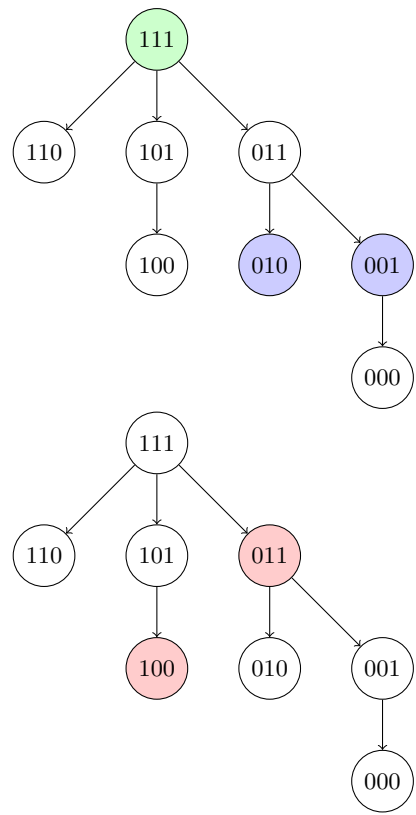
\begin{figure}[htbp]
\begin{tikzpicture}[nodes={draw, circle}, ->]
        \node[fill = green!20]{111}
                child{node {110}}
                child{node {101}
                        child{ node{100} }
                }
                child{node {011}
                        child [missing]
                        child{node[fill = blue!20]{010}}
                        child{node[fill = blue!20]{001} 
                                child{node{000}}
                        }
                };
\end{tikzpicture}
\begin{tikzpicture}[nodes={draw, circle}, ->]
        \node{111}
                child{node{110}}
                child{node {101}
                        child{ node[fill= red!20]{100} }
                }
                child{node[fill = red!20] {011}
                        child [missing]
                        child{node{010}}
                        child{node{001} 
                                child{node{000}}
                        }
                };
\end{tikzpicture}
\caption{\label{fenwick_trees} \textbf{Top:}  Update set (green) and Flip set (blue) for $i=3$ $(011)$ on Fenwick tree with $N=8$. \textbf{Bottom:} Parity set (red) for $i=5$ $(101)$ on Fenwick tree with $N=8$.}
\end{figure}

Our construction may be regarded as a particular extension of the Bravyi-Kitaev mapping. The Bravyi-Kitaev mapping stores the parity of fermionic modes in a non-local Fenwick tree data structure \cite{havlivcek2017operator}. A Fenwick tree \cite{fenwick1994new} is a partial ordering of binary representations where every node is a copy of its parent with one of the $1$s changed to a $0$ as shown in Figure~\ref{fenwick_trees}. In this section we outline how the Bravyi-Kitaev mapping encodes $m$ fermionic modes:

First, each fermionic mode is assigned a corresponding qubit and a node in a Fenwick tree. Every qubit stores the total parity of the fermionic modes below it in the Fenwick tree. This allows fermionic operators to be constructed with Pauli-weight $O(\log m)$.

These operators (as shown in~\cite{seeley2012bravyi}) are constructed using three easily calculated subsets of the Fenwick tree: the Update ($U$), Parity ($P$), and Flip ($F$)  sets.
\begin{dmath*}
a_i \rightarrow X_{U_i}  \sigma^-_i  Z_{ P_i/F_i} \rightarrow \frac{1}{2}\left( \bigotimes_{j \in U_i} X_j \otimes X_i \otimes \bigotimes_{j \in P_i} Z_j + i \bigotimes_{j \in U_i} X_j \otimes Y_i\otimes \bigotimes_{j \in P_i/F_i} Z_j \right)
\end{dmath*}
\begin{dmath*}
a^{\dagger}_i \rightarrow X_{U_i}  \sigma^+_iZ_{ P_i/F_i} \rightarrow \frac{1}{2}\left( \bigotimes_{j \in U_i} X_j \otimes X_i \otimes \bigotimes_{j \in P_i} Z_j - i \bigotimes_{j \in U_i} X_j \otimes Y_i\otimes \bigotimes_{j \in  P_i/F_i} Z_j \right) 
\end{dmath*}
Each of the above sets is defined (similarly to the original formulation \cite{seeley2012bravyi}) as follows :

Use the binary representation: $\alpha = \alpha_{l-1}\ldots \alpha_0$ and $\beta = \beta_{l-1}\ldots \beta_0$, where index $0$ denotes the most significant bit. Define $\prec$ to represent the partial ordering of the Fenwick tree. Fermionic modes are zero-indexed.

\begin{definition} Let the Update Set $U(\alpha)$ be such that $\beta \in U(\alpha)$ if and only if for some $i_0$, $\alpha_{i_0} = 0$ and $\alpha_i = \beta_i$ for $i\geq i_0$ and $\beta_i = 1$ for $i \leq i_0$. Equivalent to $\beta \in U(\alpha)$ if and only if $\alpha \prec \beta$.
\end{definition}

\begin{definition} Let the Parity set $P(\alpha)$ contain elements $\beta \in P(\alpha)$ if and only if for some $i_0$: $\alpha_{i_0} = 1$, $\beta_{i_0} = 0$, $\beta_i = \alpha_i$ for $i> i_0$ and $\beta_i = 1$ for $i< i_0$, then $
        \sum_{j \in P(i)} q_j = \sum_{j < i} f_j$
\end{definition}

\begin{definition} Let the Flip set $F(\alpha)$ contain elements $\beta \in F(\alpha)$ if and only if for some $i_0$: $\beta_{i_0} = 0$, $\beta_i = \alpha_i$ for $i\neq i_0$ and $\alpha_i = 1$ for $i\leq i_0$, then $
 \sum_{j \in F(i)} q_j = \sum_{j \prec i} f_j$
\end{definition}

Examples of these sets are given in Figure~\ref{fenwick_trees}. We present three simple algorithms for generating these sets which will later be used in our construction: \hyperref[GeonerateUpdateSet]{\textsc{GenerateUpdateSet}}, \hyperref[GenerateParitySet]{\textsc{GenerateParitySet}}, and \hyperref[GenerateFlipSet]{\textsc{GenerateFlipSet}}.
 \noindent

\begin{algorithm}[htbp]
\caption{\textsc{GenerateUpdateSet}\label{GenerateUpdateSet}}
\begin{algorithmic}[1]
\Require{Fermionic enumeration: $\alpha$, Number of fermionic modes: $n$}
\Ensure{Update Set $U_{\alpha}$}
\State updateSet = \{\}
\State $l \gets $ length of binary representation of $\alpha$
\State $\beta \gets \alpha$
\For{$i$ counting from $0$ to $l-1$}
    \If{$\alpha_i$ is $0$}
        \State $\beta \gets \beta_{l-1}\ldots\beta_{i+1}1\beta_{i-1}\ldots\beta_0$
        \If{$\beta < n$}
            \State Add $\beta$ to updateSet
        \EndIf
    \EndIf
\EndFor

\end{algorithmic}
\end{algorithm}

\begin{algorithm}[htbp]
    \caption{\label{GenerateParitySet}\textsc{GenerateParitySet}}
\begin{algorithmic}[1]
\Require{Fermionic enumeration: $\alpha$}
\Ensure{Parity Set $P_{\alpha}$}
\State paritySet = \{\}
\State $l \gets $ length of binary representation of $\alpha$
\For{$i$ counting down from $l-1$ to $0$}
    \If{$\alpha_i$ is $1$}
        \State $\beta \gets \alpha_{l-1}\alpha_{l-2}\ldots\alpha_{i+1}01\ldots 1$
        \State Add $\beta$ to paritySet 
    \EndIf
\EndFor
\end{algorithmic}
\end{algorithm}
\begin{algorithm}[htbp]
    \caption{\label{GenerateFlipSet}\textsc{GenerateFlipSet}}
\begin{algorithmic}[1]
\Require{Fermionic enumeration: $\alpha$}
\Ensure{Flip Set $F_{\alpha}$}
\State flipSet = \{\}
\State $l \gets $ length of binary representation of $\alpha$
\State $i \gets 0$
\While{$\alpha_i$ is $1$}
    \State $\beta \gets \alpha_{l-1}\ldots \alpha_{i+1}0\alpha_{i-1} \ldots \alpha_0$
    \State Add $\beta$ to flipSet
    \State $i \gets i + 1$
\EndWhile
\end{algorithmic}
\end{algorithm}
\pagebreak
\subsection{Auxiliary Qubit Mapping}
\label{section:auxillaryMappings}
Here we recall the formalization of the auxiliary mappings from~\cite{steudtner2019quantum}. The principle behind auxiliary mapping schemes is to create stabilisers which are similar to non-local interaction terms in the Hamiltonian, so they can be used to cancel each other out. For example, consider an $n$-qubit Hamiltonian consisting of $m$ Pauli-strings $h_i$ some of which contain a non-local Pauli-string $p$:
\begin{equation}
    H = \sum_{i=0}^m h_i
\end{equation}
\begin{itemize}
\item Let $\ket{\psi}_S$ be an arbitrary state on the $n$-qubit system ($\mathcal{H}_S$)
\item Introduce an ancillary qubit $\ket{0}_A$ ($\mathcal{H}_A$)
\item Entangle the ancilla with the $n$-qubit state by applying a given unitary $V$
\begin{equation}
    V \ket{\psi}_S \ket{0}_A = \ket{\tilde \psi}_{SA}
\end{equation}
\item Find $\sigma$ such that $(p \otimes \sigma)$ is a stabiliser:
\begin{equation}
    (p \otimes \sigma) \ket{\tilde \psi}_{SA} = \ket{\tilde \psi}_{SA}
\end{equation}
\item Modify $H$ so it commutes with the stabiliser (\ref{eq:commute_with_hamiltonian}) and thus acting with the modified Hamiltonian $\tilde H$ on a stabiliser state always produces another stabiliser state. This involves adding an operator $\kappa$ to each term $h_i$ in the Hamiltonian that anti-commutes with $(p \otimes \sigma)$ to produce a commuting term $h_i^{(\kappa)} = h_i \otimes \kappa$. This modification must be such that its action on the entangled system is equivalent to the action of the original Hamiltonian on the $n$-qubit state (\ref{eq:equivalent_to_original_hamiltonian}).
\begin{equation}
\label{eq:commute_with_hamiltonian}
    H^{(\kappa)}(p \otimes \sigma)  \ket{\tilde \psi}_{SA} =(p \otimes \sigma)  H^{(\kappa)}\ket{\tilde \psi}_{SA}  
\end{equation}
\begin{equation}
    \label{eq:equivalent_to_original_hamiltonian}
    H^{(\kappa)}\ket{\tilde \psi}_{SA} = V (H \otimes I) \ket{\psi}_S \ket{0}_A
\end{equation}
\item Now we can use the stabiliser to cancel out the non-local $p$ string in the Hamiltonian to get a simplified Hamiltonian $\tilde H$ which is logically identical.
\begin{equation}
   \tilde h_i = \begin{cases} h_i^{(\kappa)} (p \otimes \sigma)  &  p \in h_i \\ 
   h_i^{(\kappa)} & p \notin h_i \end{cases}
\end{equation}
\begin{equation}
    \tilde H = \sum_{i=0}^m \tilde h_i
\end{equation}
\end{itemize}
To perform this procedure to remove multiple non-local strings $p_i$, we can simply repeat the process with additional ancillas. However as the stabilisers $p_i \otimes \sigma_i$ must share common stabiliser states, it is only possible to simultaneously remove commuting strings. 
\subsection{Steiner Tree Solver}
\label{section:SCIPJACK}
In our analysis we evaluate the performance of different mappings on a square-lattice qubit architecture. Some mappings require 2-qubit gates to be performed on non-neighbouring qubits and so we need additional SWAP gates. Calculating the optimum choice of SWAP gates requires finding the minimum path length required to connect all the qubits involved in a given fermionic operation. This is an example of an NP-hard problem known as the Steiner Tree problem:

\begin{algorithm}[h]
\centering
\caption*{\label{alg:steinerTree}\textsc{Steiner Tree Problem}}
\begin{algorithmic}
\State \textsc{Given an undirected connected graph $G = (V,E)$ and a set $T \in V$ of terminals, find a tree $S \in G$ of minimum length that spans $T$}
\end{algorithmic}
\end{algorithm}

In our case the set of terminals corresponds to the set of qubits involved in the given fermionic operation which we want to find the minimum path of SWAP gates between. In our calculations we use a top-ranking Steiner Tree solver SCIP-JACK \cite{rehfeldt2021implications} to find this minimum path.  SCIP-JACK consists of three component parts: preprocessing, primal heuristics, and a branch-and-cut procedure.

The preprocessing uses reduction techniques to reduce the size of the graph whilst still allowing at least one solution to the reduced problem to be converted into an optimal solution to the original graph. Examples of these techniques include merging together connected vertices that share no other neighbours or discarding 1-degree non-terminal vertices.

The primal heuristic finds good or optimal solutions to the problem. In SCIP-JACK the one-phase repetitive shortest path heuristic \cite{aragao2002implementation} is used. This starts from a vertex and uses Dijkstra’s algorithm \cite{dijkstra2022note} to find the shortest path to the nearest terminal. This procedure is then reiterated with the modification to Dijkstra’s algorithm that the tentative distance assigned to each vertex is carried over. Once this has built up a subgraph containing all the terminals, a minimum spanning tree is constructed covering the subgraph, and any 1-degree non-terminal vertices are pruned.

The branch-and-cut procedure is the backbone of the program used to compute a lower bound and prove the optimality of the solution. First, the problem is expressed in terms of a integer programming problem by considering $x_e$ to be binary variables representing whether edge $e \in E$ is included in the Steiner tree:

\begin{algorithm}
\caption*{\label{alg:steinerTreeCut}\textsc{Steiner Tree Problem Undirected Cut Formulation}}
\begin{algorithmic}[1]
\Statex Define $\delta (U) =\{\{u,v\} \in E: u \in U, v \in V/U \}$
\Statex
\Statex Find min $\sum_e x_e$ under the following constraints:
\State $x(\delta (W)) \geq 1$ for all $W \in V$, $0 < |W \cap T| < |T|$
\State $x_e \in \{0,1\} \forall e \in E$
\end{algorithmic}
\end{algorithm}

Linear programming is used to initially optimise the solution when considering only a subset of these constraints providing a lower bound on an optimal solution. The primary heuristic is then run with edge costs given by $1-x_e$ to find a feasible solution (reasonably similar to the LP solution) providing an upper bound on the optimal solution. Then, the algorithm considers the restraints ("cuts") that the current solution violates and adds them to the subset of constraints in the linear programming model. This model is repeatedly solved and expanded until the subset of constraints has grown sufficiently. At this point, if the solution is integral then we have found an optimal solution, otherwise the algorithm "branches" by creating two new sub-problems (nodes) each with an additional constraint on one of the variables ($x_e$) that forces it to be either 0 or 1 in each problem respectively. The process is then repeated again on each of these nodes (potentially with more constraints added). If for any node the optimal solution exceeds any of the heuristic solutions found, the node is discarded (pruned).

\section{Novel Fermion-to-qubit mappings}
We present two novel mapping algorithms: Hybrid mapping (\hyperref[section:Hybridmapping]{A}) and Hybrid\+ mapping (\hyperref[section:auxHybridmapping]{B}). The Hybrid mapping combines the low connectivity constraints of the Jordan-Wigner mapping with the increased locality of the Bravyi-Kitaev mapping. The Hybrid\+ mapping builds on the strengths of the Hybrid mapping and expands the mapping to include aspects of AQM \cite{steudtner2019quantum} to increase the locality using ancilla qubits.
In this section we outline their construction in detail, and then in Section~\ref{section:algorithmanalysis} we will analyse their performance. 
Both these mappings have been designed for a square-lattice fermionic interaction graph. For the sake of clarity, they are both described with $n \times n$ cells, however, it is trivial to extend them to rectangular $l \times h$ cells. It is also possible to perform these mappings with irregular varying cell sizes.
\subsection{Hybrid mapping algorithm}
\label{section:Hybridmapping}
\begin{figure}[hbtp] 
        \centerline{
\begin{tikzpicture}[nodes={draw, circle}, ->]
        \node (3) at (0,1) {3};
        \node (1) at (1,1) {1};
        \node (0) at (1,0) {0};
        \node (2) at (0,0) {2};
        \draw [->] (3) -- (1);
        \draw [->] (3) -- (2);
        \draw [->] (1) -- (0);
        \node[fill = green!20] (7) at (2,1) {7};
        \node[fill = red!20]  (5) at (3,1) {5};
        \node (4) at (3,0) {4};
        \node[fill = orange] (6) at (2,0) {6};
        \draw [->] (7) -- (5);
        \draw [->] (7) -- (6);
        \draw [->] (5) -- (4);
        \node (11) at (0,-1) {31};
        \node (9) at (1,-1) {29};
        \node (8) at (1,-2) {28};
        \node (10) at (0,-2) {30};
        \draw [->] (11) -- (9);
        \draw [->] (11) -- (10);
        \draw [->] (9) -- (8);
        \node[fill = orange] (27) at (2,-1) {27};
        \node[fill = blue!20]  (13) at (3,-1) {25};
        \node (12) at (3,-2) {24};
        \node[fill = blue!20]  (14) at (2,-2) {26};
        \draw [->] (27) -- (13);
        \draw [->] (27) -- (14);
        \draw [->] (13) -- (12);
        
        \node[fill = red!20] (11) at (4,1) {11};
        \node (9) at (5,1) {9};
        \node (8) at (5,0) {8};
        \node (10) at (4,0) {10};
        \draw [->] (11) -- (9);
        \draw [->] (11) -- (10);
        \draw [->] (9) -- (8);
        \node[fill = red!20] (15) at (6,1) {15};
        \node (13) at (7,1) {13};
        \node (12) at (7,0) {12};
        \node (14) at (6,0) {14};
        \draw [->] (15) -- (13);
        \draw [->] (15) -- (14);
        \draw [->] (13) -- (12);
        \node[fill = red!20]  (23) at (4,-1) {23};
        \node (9) at (5,-1) {21};
        \node (8) at (5,-2) {20};
        \node (26) at (4,-2) {22};
        \draw [->] (23) -- (9);
        \draw [->] (23) -- (26);
        \draw [->] (9) -- (8);
        \node[fill = red!20]  (19) at (6,-1) {19};
        \node (13) at (7,-1) {17};
        \node (12) at (7,-2) {16};
        \node (14) at (6,-2) {18};
        \draw [->] (19) -- (13);
        \draw [->] (19) -- (14);
        \draw [->] (13) -- (12);
        \draw [-, orange] (6) -- (27);
\end{tikzpicture}}
\caption{\label{fig:fenwick_tree_cells} Hybrid mapping with $2\times 2$ cells showing the different sets of qubits involved in a fermionic operation between mode 6 and 27. Orange = targets, Blue = Flip set, Red = Parity set, Green = Update set}
\end{figure}
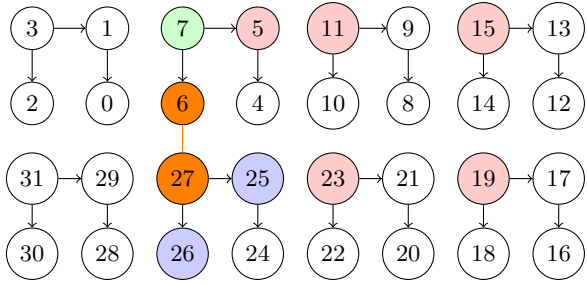

\noindent
\textbf{Input:} \textit{$N\times N$ Fermion Square-Lattice Interaction Graph, Cell Shape $n \times n$}, $1\le n\le N$ and $n$ is a factor of $N$\\
\textbf{Output:} \textit{Pauli-strings corresponding to interaction edges, Enumerated square-lattice of qubits}\\
\textbf{Algorithm:}
\begin{enumerate}
    \item Divide the square lattice of fermions into cells of size $n \times n$
    \item Assign a \textit{cell number} $(c_i)$ to each of these cells in a Z pattern starting with the top-left most cell (illustrated by Figure~\ref{ZPattern} in the Appendix). 
    This enumeration reflects the Jordan-Wigner-esque properties of the mapping.
    \item Within these cells assign each fermion a \textit{mode number} $(f_i)$ from $1$ to $n\times n$ in a Z pattern starting with the top-left most fermion. We can now refer to the $i$-th fermion in the graph via the coordinate $(c_i, f_i)$.
    \item Repeat this procedure on a square qubit lattice of the same size so for every fermion there is an equivalent qubit labeled with the same coordinate
    \item For each interaction edge in your fermion interaction graph between $(c_1, f_1)$ and $(c_2, f_2)$ generate the Pauli-string representation:
    \begin{enumerate}
        \item Generate the update $U_i$, flip $F_i$, and parity $P_i$ sets for $f_1$ and $f_2$ independently in a Fenwick tree of size $n \times n$\\
        \textsc{$U_i = (c_i, $ \hyperref[GenerateUpdateSet]{GenerateUpdateSet}($f_i, n\times n$))}\\
        \textsc{$F_i = (c_i, $ \hyperref[GenerateFlipSet]{GenerateFlipSet}($f_i$))}\\
        \textsc{$P_i = (c_i, $ \hyperref[GenerateParitySet]{GenerateParitySet}($f_i$))}
        \item Add the roots of every cell lower in the Z pattern than $c_1$ and $c_2$ to their respective parity sets. \\
        \textsc{For all $c_j < c_1$ add $(c_j, n \times n)$ to $P_1$}\\
        \textsc{For all $c_j < c_2$ add $(c_j, n \times n)$ to $P_2$}
        \item Pauli-string representation is given by:\\
            \begin{dmath*}[style={\footnotesize}]
                    a_1a^{\dagger}_2  \rightarrow X_{U_1}  \sigma^-_1  Z_{P_1/F_1}  X_{U_2}  \sigma^+_2Z_{P_2/F_2}
                    = \frac{1}{4}\left( \bigotimes_{j \in U_1} X_j \otimes X_{(c_1, f_1)} \otimes \bigotimes_{j \in P_1} Z_j + i \bigotimes_{j \in U_1} X_j \otimes Y_{(c_1, f_1)}\otimes \bigotimes_{j \in P_1/F_1} Z_j \right)\\ \left( \bigotimes_{j \in U_2} X_j \otimes X_{(c_2, f_2)} \otimes \bigotimes_{j \in P_2} Z_j - i \bigotimes_{j \in U_2} X_j \otimes Y_{(c_2, f_2)}\otimes \bigotimes_{j \in P_2/F_2} Z_j \right)\\ 
                    = \frac{1}{4}\bigotimes_{j \in (U_1 \oplus U_2)} X_j \bigotimes_{j \in (P_1/F_1 \oplus P_2/F_2)} Z_j \otimes \\\left( X_{(c_1, f_1)} \otimes \bigotimes_{j \in F_1} Z_j + i Y_{(c_1, f_1)} \right) \left( X_{(c_2, f_2)} \otimes \bigotimes_{j \in F_2} Z_j - i  Y_{(c_2, f_2)} \right) 
            \end{dmath*}    
    \end{enumerate}
\end{enumerate}
This mapping effectively splits up the square lattice into a series of Fenwick trees (Bravyi-Kitaev maps) and connects them via a Jordan-Wigner mapping on the roots, as demonstrated in Figure~\ref{fig:fenwick_tree_cells}.
\subsection{Hybrid\+ mapping algorithm}
\label{section:auxHybridmapping}
\noindent
\textbf{Input:} \textit{$N\times N$ Fermion Square-Lattice Interaction Graph, Cell Shape $n \times n$}, $n \times n = 2^k$ for some integer $k$, $n$ is a factor of $N$\\
\textbf{Output:} \textit{Pauli-strings corresponding to interaction edges, Enumerated square-lattice of qubits plus required ancillas, Preprocessing entanglement gates}\\
\textbf{Algorithm:}
\begin{enumerate}
    \item Perform steps 1-4 of the Hybrid mapping algorithm, but additionally assign the \textit{cell number} ($c_i$) in a S pattern (illustrated by Figure~\ref{SPattern} in Appendix~\ref{appendix}).
    \item For every cell $c_i$, add one ancilla to the qubit enumeration scheme. This ancilla qubit is labelled $(c_i, n \times n +1)$ and only needs to be connected to the root of the cell $(c_i, n\times n)$ and the ancillas of the neighbouring cells.
    \item Prepare a stabiliser for each ancilla containing the string of Z gates involved in performing a fermionic operation between the roots of the current cell and the cell vertically above.
    (A detailed description of these stabilisers is given in Appendix~\ref{appendix} Figure~\ref{fig:stabilisers}). The stabilisers are created using the method described in \cite{steudtner2019quantum} by applying the following series of entangling gates: 
    \begin{enumerate}
        \item Apply a Hadamard gate to every ancilla 
        \item Entangle the edges of the rows by applying either of the gates controlled by the ancilla C shown in Figure~\ref{fig:entangleEdge} depending on the direction of the S-pattern on the row.
        \begin{figure}
        \includegraphics[page=4, width=0.5\textwidth  , height=6cm]{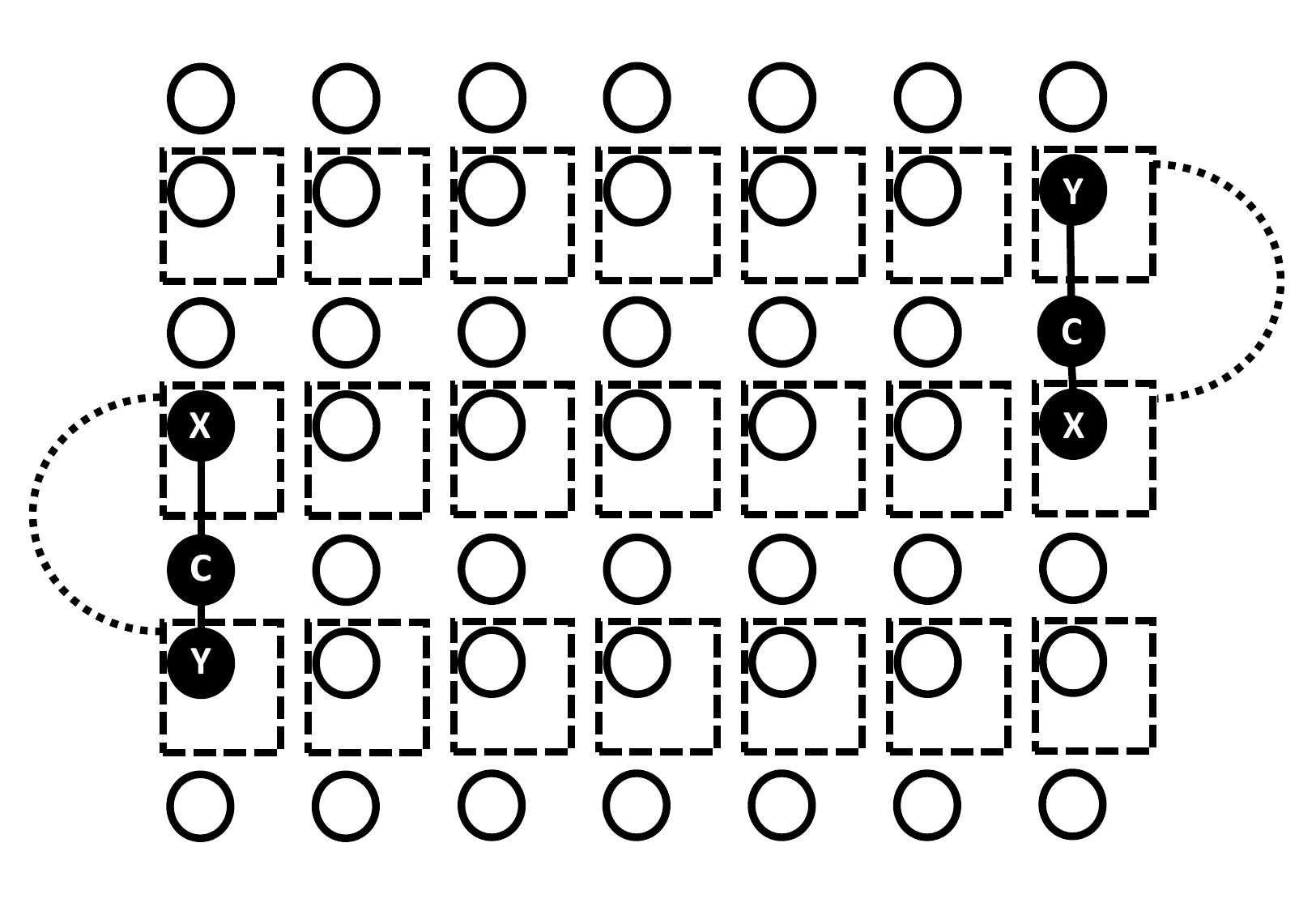}
        \caption{\label{fig:entangleEdge}Gates performed on edges of rows conditional upon the ancillas at the end of each row labeled with $C$.  The dashed boxes represent the cells, whilst the circles inside them represent the roots of each cell, and the circles without boxes represent the ancillas.}
    \end{figure}
    \item Move along each row applying either of the gates controlled by ancilla C shown in Figure~\ref{fig:entangleRow} depending on the direction of the S-pattern on the row.
   \begin{figure}
        \includegraphics[page=5, width=0.5\textwidth  , height=6cm]{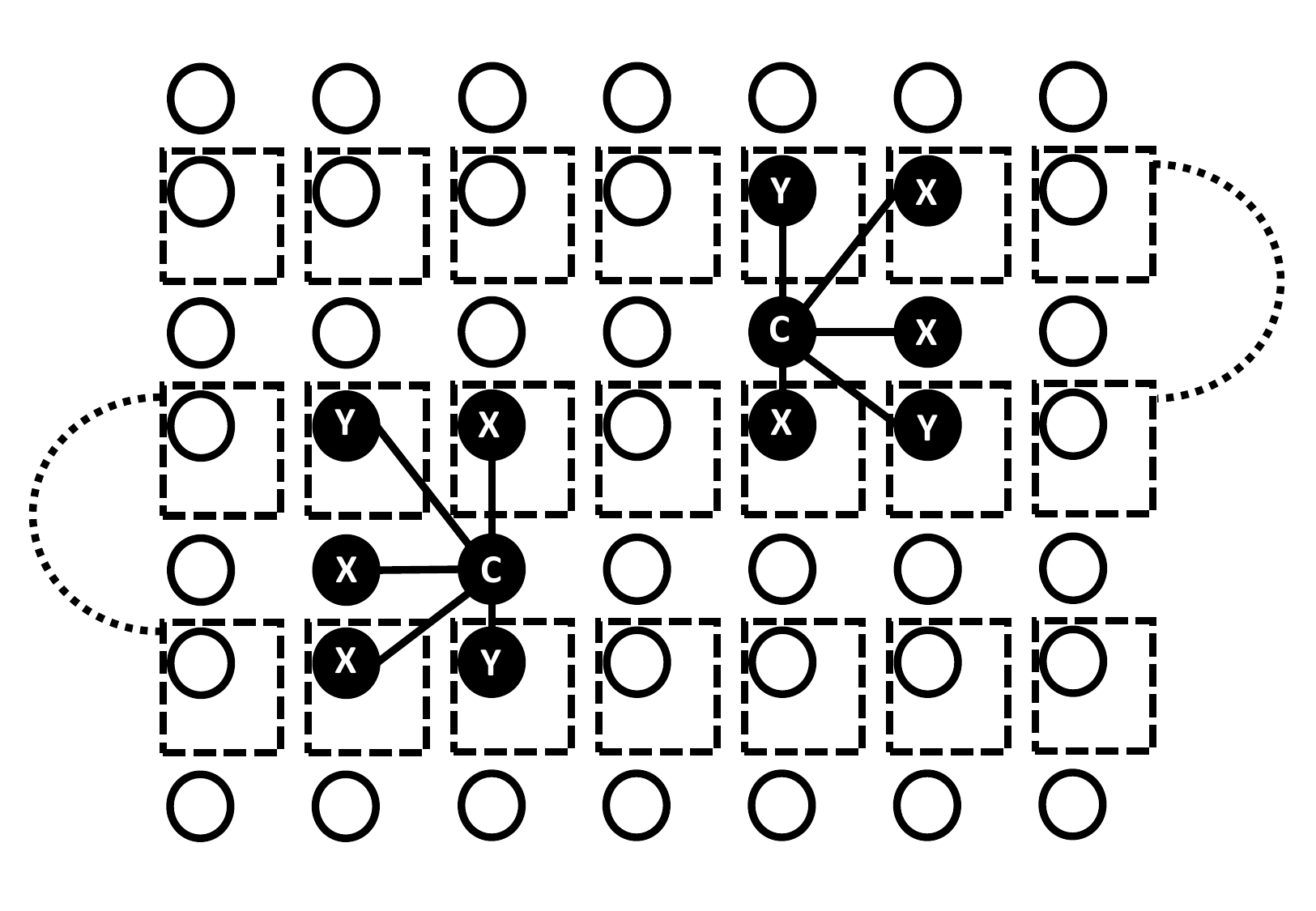}
        \caption{\label{fig:entangleRow}Examples from the two (for odd/even rows) series of gates applied conditional upon the ancillas labeled with $C$. Each element in the series is identical to the previous but shifted one row to the left (odd rows) or right (even rows). The first element in each series is the 2nd rightmost (odd rows) and 2nd leftmost (even rows).}
    \end{figure}
    \\\\This technique exploits the newly created stabiliser of the previously entangled ancilla to cancel out most of the gates required to entangle the next ancilla in the row.
    \end{enumerate}
    \item For each interaction edge in your fermion interaction graph between $(c_1, f_1)$ and $(c_2, f_2)$ generate the Pauli-string representation:
    \begin{enumerate}
        \item Perform steps 5a)-c) of the Hybrid mapping algorithm
        \item We will only consider physical (even) fermionic operations, i.e. those which preserve the fermion number parity. For example:
        \begin{dmath*}
                    a_1a^{\dagger}_2 +a_2a^{\dagger}_1  \rightarrow X_{U_1}  \sigma^-_1  Z_{P_1/F_1}  X_{U_2}  \sigma^+_2Z_{P_2/F_2} +  X_{U_2}  \sigma^-_2  Z_{P_2/F_2}  X_{U_1}  \sigma^+_1Z_{P_1/F_1}
                    \rightarrow \frac{1}{2}\bigotimes_{j \in (U_1 \oplus U_2)} X_j \bigotimes_{k \in (P_1/F_1 \oplus P_2/F_2)} Z_k \> \\ \left( X_{(c_1, f_1)} \otimes X_{(c_2, f_2)} \bigotimes_{k \in F_1} Z_k \bigotimes_{j \in F_2} Z_j  + Y_{(c_1, f_1)}  \otimes Y_{(c_2, f_2)} \right) 
            \end{dmath*}
        In what follows we only need to consider how these operators act on the roots of the cells $(c_i, n\times n)$, as the stabiliser only deals with the cell roots. As $n\times n = 2^k$ each cell will contain a connected Fenwick tree, so each root will either be in the update set or be one of the targets. Therefore, each root is either acted on by $X_{(c_i, f_i)} $ or $X_{(c_t, f_t)} \otimes X_{(c_i, f_i)} \bigotimes_{k \in F_t} Z_k \bigotimes_{j \in F_i} Z_j  + Y_{(c_t, f_t)}  \otimes Y_{(c_i, f_i)} $. 
        \item We need to modify every string so they preserve the stabilisers. For each stabiliser, add a Z gate (acting on the associated ancilla) to every string that anti-commutes with the stabiliser. We are also able to shorten long S pattern Pauli strings by multiplying by appropriate stabilisers. Figure~\ref{fig:verticalInteractionXodd} demonstrates how the Hybrid\+ mapping modifies a Pauli-string representing a local fermionic operation in the Hybrid mapping scheme. Further examples are illustrated by Figures~\ref{fig:verticalInteractionYodd} and \ref{fig:horizontalInteractionXYYX} in Appendix~\ref{appendix}.\\\\
        If neither of the target fermions are the roots of their cell then the fermionic operation (on the level of cell roots) will simply be $XZ \ldots ZX$. However, if one of them is a root it will also include terms such as $XZ\ldots ZY$ and $YZ \ldots ZX$ (and potentially $YZ\ldots ZY$ if they are both roots).
   \begin{figure}[hbpt]
        \includegraphics[page=6, width=0.5\textwidth , height=6cm]{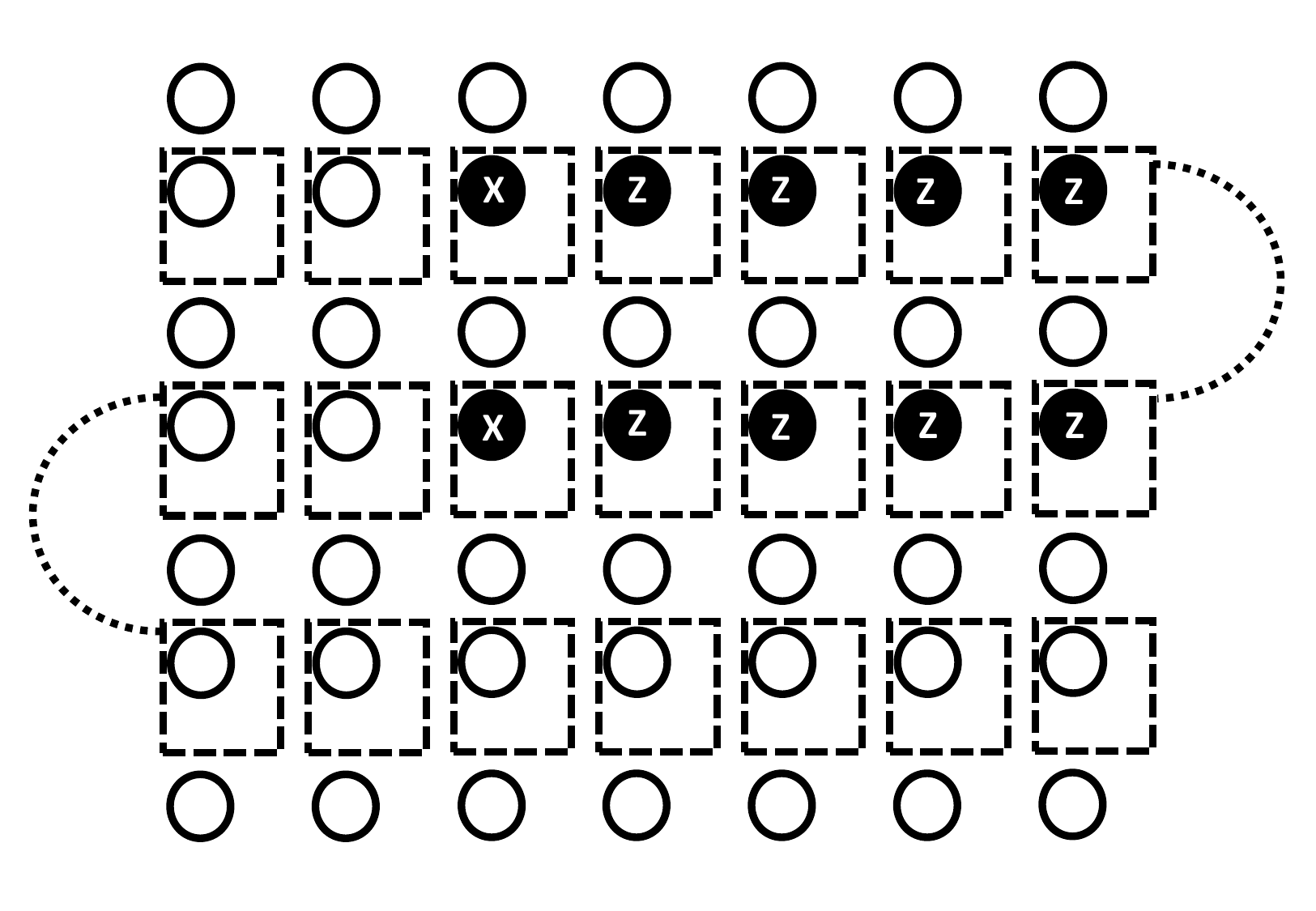}
        \tikz[baseline=-\baselineskip]\draw[ultra thick,->] (0,0) -- ++ (0,-1);
        \includegraphics[page=7, width=0.5\textwidth , height=6cm]{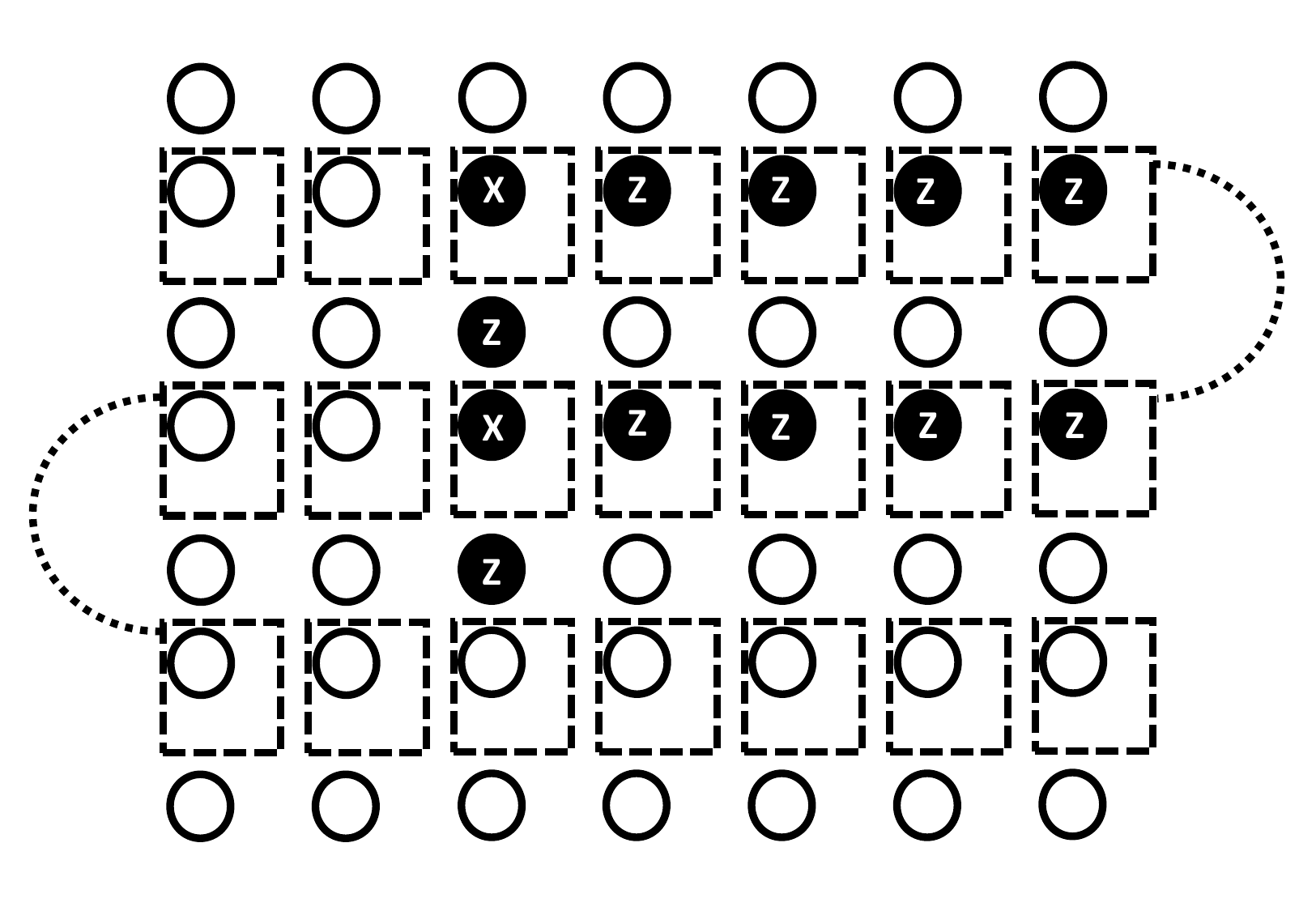}
        \tikz[baseline=-\baselineskip]\draw[ultra thick,->] (0,0) -- ++ (0,-1);
        \includegraphics[page=8, width=0.5\textwidth  , height=6cm]{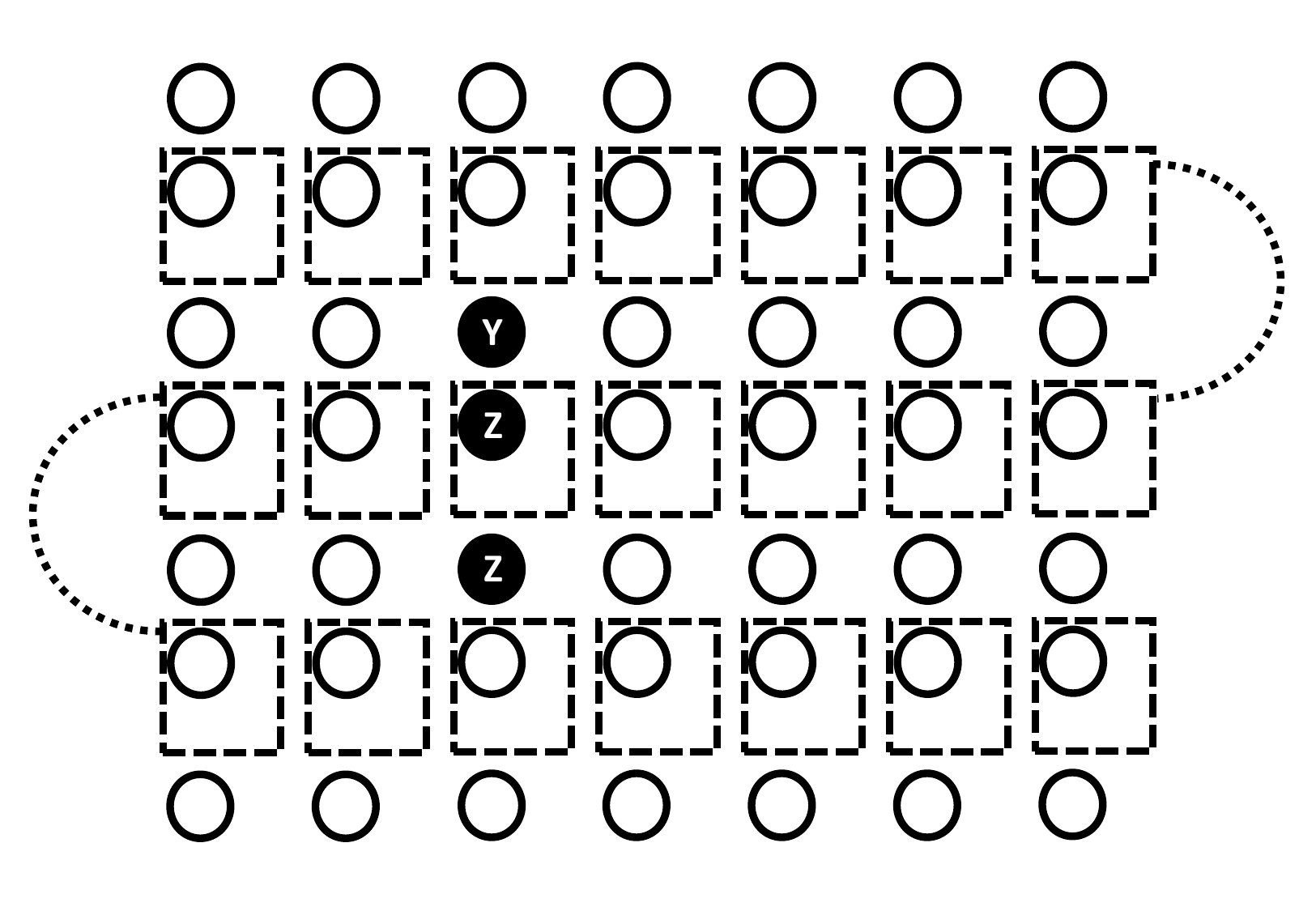}
        \caption{\label{fig:verticalInteractionXodd}Diagram showing how the Pauli-string representing a vertical hopping term (top) is modified by the Hybrid\+ mapping algorithm. First, Z gates are added to all the ancillas with associated stabilisers that anti-commute with the string (middle). Then, a stabiliser shown in Figure~\ref{fig:stabilisers} is applied canceling out a large portion of the string to give the final local operation (bottom)}
    \end{figure}
    \end{enumerate}
\end{enumerate}
 This algorithm essentially extends the Jordan-Wigner portion of the Hybrid mapping scheme in the same way AQM extends the Jordan-Wigner mapping.
\section{Algorithm analysis}
\label{section:algorithmanalysis}
\subsection{Theoretical Average Pauli-weight of $n\times n$ Hybrid on an $N\times N$ lattice}
\label{section:hybridanalysis}
There are two primary contributions to the Pauli-weight of a fermionic operation in the Hybrid scheme:
\begin{enumerate}
    \item Local (Bravyi-Kitaev type) gates acting on qubits within the target cells
    \item Non-local (Jordan-Wigner type) gates acting on the roots of the cells between the two target cells in the Z-pattern
\end{enumerate}
The first contribution scales with $O(\log(n^2))$, whereas the second contribution scales with $O(N)$.

We can analytically calculate the second contribution:
\begin{itemize}
    \item Any interactions within a single cell make no contributions here
    \item Horizontal interaction terms between cells only contribute the root of one the target cells. However, as this root will already either be the target or in the update set we can ignore this contribution.
    \item Vertical interaction terms between qubits in different cells contribute $\frac{N}{n} -1$. Therefore, the total contributions from each row is:
    \begin{equation}
            \sum_{i=1}^{N} \frac{N}{n} -1 = N\left(\frac{N}{n} -1\right)
    \end{equation}
    Therefore, as there are $\frac{N}{n}$ rows the total of the second contribution is:
    \begin{equation}
        \frac{N^2}{n}\left(\frac{N}{n}-1\right)
    \end{equation}
    As there are $2N(N - 1)$ interactions in the whole lattice this gives the following contribution to the average Pauli-weight:
    \begin{equation}
        \frac{N}{2(N-1)n}\left(\frac{N}{n}-1\right)
    \end{equation}
\end{itemize}
Therefore, the average Pauli-weight of $n\times n$ Hybrid on an $N\times N$ lattice is:
\begin{equation}
        \frac{N(N-n)}{2(N-1)n^2} + O(\log n) 
\end{equation}
and so for large lattices:
\begin{equation}
        \frac{N}{2n^2} + O(\log n) 
\end{equation}
When $N\gg n$, the Hybrid mapping produces $n^2$ times smaller Pauli-strings than the Jordan-Wigner mapping.
\subsection{Theoretical Average Pauli-weight of $n\times n$ Hybrid\+ on an $N\times N$ lattice}
Similarly to the Hybrid mapping, there are two primary contributions to the Pauli-weight of a fermionic operation:
\begin{enumerate}
    \item Local (Bravyi-Kitaev type) gates acting on qubits within the target cells
    \item Local (AQM type) gates acting on ancilla qubits
\end{enumerate}
The first contribution scales with $O(\log(n^2))$, whereas the second contribution scales with $O(\frac{1}{n})$ (as the ratio of inter-cell/intra-cell interactions decreases as $n$ increases). Therefore, as $N$ (and therefore the number of fermionic modes) increases the average Pauli-weight will remain constant.

\subsection{Varying impact of connectivity}
\label{section:connectivity}
The limitations of the underlying qubit architecture has a large impact on the performance of various mappings. In this section, we demonstrate the advantages our mappings give over existing schemes when used on a quantum computer with the common square-lattice connectivity. The connectivity constraints impact mappings by requiring SWAP gates to perform entangled operations between physically non-local qubits. In our analysis, we compare the number of qubits acted upon by a Pauli or SWAP gate during a fermionic operation (the \textit{interaction qubit count}). Below we consider the varying impact of connectivity upon different mappings.

\textbf {Jordan-Wigner Mapping}. The Jordan-Wigner mapping only requires a linear connectivity mapping, so it requires no additional gates to use on a square-lattice of qubits. However, it has been shown that an optimal choice of enumeration can give a constant improvement in the Pauli weight~\cite{chiew2021optimal}.

\textbf {Bravyi-Kitaev Mapping}.
The Bravyi-Kitaev mapping requires a very high level of connectivity. In order to avoid any additional gates one must be able to embed an $\log N$ degree graph (the Fenwick tree) within the qubit architecture. As a square-lattice of connectivity has degree $4$ it is not possible to embed a Bravyi-Kitaev mapping of more than 16 fermionic modes. Therefore, for Bravyi-Kitaev mappings on larger systems neighbouring Fenwick tree nodes will not be connected, so additional SWAP gates are required to perform entangled gates on these qubits.

We found that simulating these additional SWAP gates drastically reduced the performance of the Bravyi-Kitaev mapping. This is theoretically supported by considering the scaling of the largest interaction qubit count interactions. Consider the fermionic operation between the fermions corresponding to qubits which are physically located at opposite ends of a row (or column), this operation must include these qubits as they are the targets. To connect just these two qubits at least $N$ qubits must be involved in the operation. Our unoptimized Z-pattern enumeration for the Bravyi-Kitaev mapping includes non-local interactions such as these, and hence the average interaction qubits scale with $O(N)$. It is feasible that an optimal Bravyi-Kitaev enumeration could be found that scales more favourably. However, this is a very hard problem as the number of possible enumeration schemes grows factorially with lattice size and there is no known solution.

\textbf {Hybrid mapping}.
The main strength of this mapping is its performance in the real-world scenario of a square-lattice connectivity. It combines the Jordan-Wigner mappings connectivity flexibility with the Bravyi-Kitaev mappings increased locality. The average interaction qubit count contribution from within the Bravyi-Kitaev cells scales as above with $O(n)$. While the Pauli operations on the roots of the cells along the S-pattern are joined up with SWAP gates making the contribution per row from vertical interactions:
\begin{equation}
        \sum_{i=1}^{N} (N-1) = N(N-1)
\end{equation}
Therefore, as there are $\frac{N}{n}$ rows and $2N(N-1)$ interactions in the whole lattice, the contribution to the average interaction qubit count is:
\begin{equation}
     \frac{N}{2n}
\end{equation}
Therefore, the average interaction qubit count scales with $O(n + \frac{N}{2n})$. So for $N \gg n$, this mapping requires $n$ times less interaction qubits than both the Bravyi-Kitaev and Jordan-Wigner mappings.

\textbf {Hybrid\+ mapping}.
This algorithm requires a slightly higher level of connectivity. Our formulation has a second square-lattice of ancilla qubits connected to the original at the roots of the cells. It could also be formulated by instead choosing one qubit in each cell of the original square lattice to be an ancilla. However, we chose our formulation as it produces a lower interaction qubit count, is simpler to consider, and still is a realistic qubit architecture.

The Hybrid\+ mapping retains a constant average Pauli-weight when considering the limitations of connectivity. This is due to the only rising contribution coming from the Bravyi-Kitaev mapping within the target cells which should scale with roughly $O(n)$, and for a given mapping $n$ is constant.
\begin{figure}[hbpt]
    \centering
    \includegraphics[width = 0.5\textwidth]{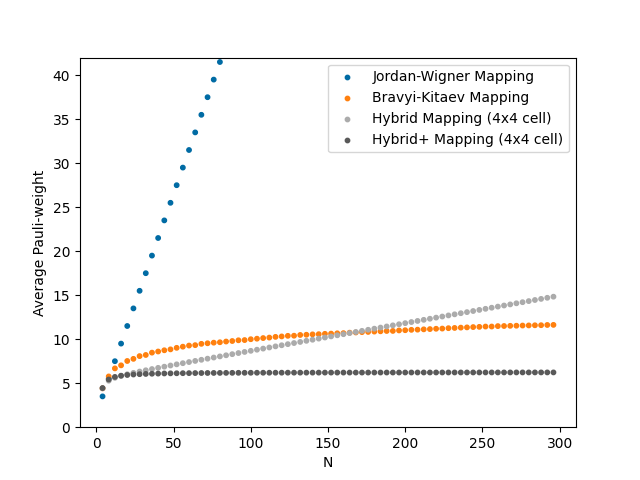}
    \caption{The average Pauli-weight of fermionic operations on an $N\times N$ lattice against $N$ when using a given mapping.}
    \label{fig:compare_mappings_ex_SWAP}
\end{figure}
\begin{figure}[hbpt]
    \centering
    \includegraphics[width = 0.5\textwidth]{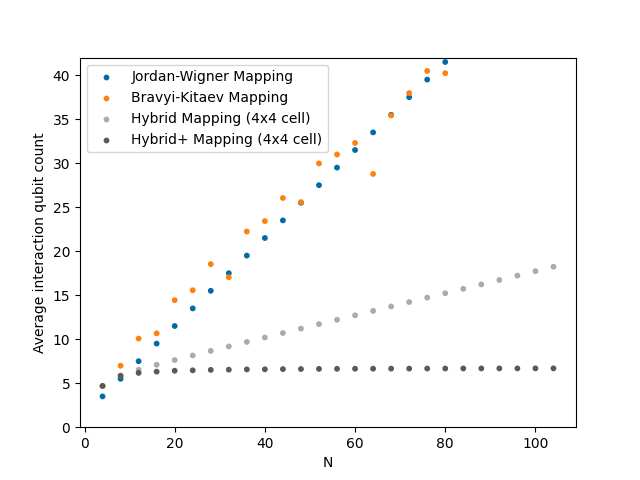}
     \caption{The average number of qubits involved in each fermionic operation on an $N\times N$ lattice against $N$ when using a given mapping in the context of a square-lattice qubit connectivity.}
    \label{fig:compare_mappings}
\end{figure}
\begin{figure}[hbpt]
    \centering
    \includegraphics[width = 0.5\textwidth]{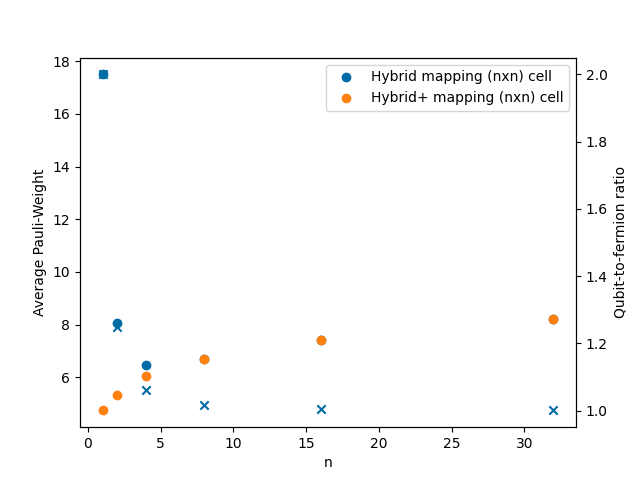}
    \caption{The average Pauli-weight of a fermionic operation on an $32\times 32$ lattice when using an $n\times n$ Hybrid/Hybrid\+ mapping against $n$. The crosses denote the qubit-to-fermion ratio of the $n\times n$ Hybrid\+ mapping for each choice of $n$.}
    \label{fig:vary_cell_sizes_ex_SWAP}
\end{figure}
\begin{figure}[hbpt]
    \centering
    \includegraphics[width = 0.5\textwidth]{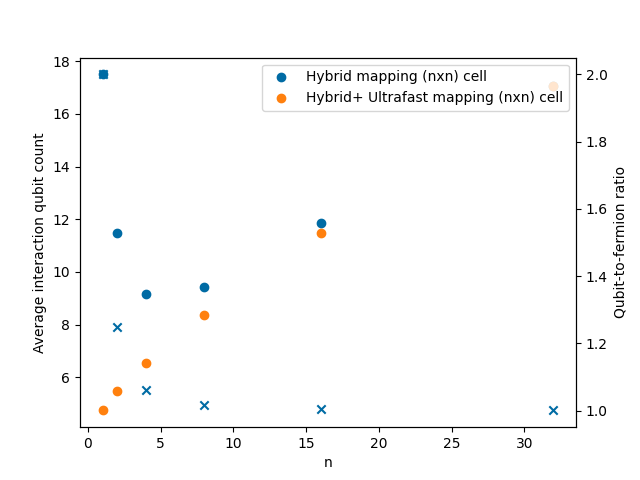}
    \caption{The average number of qubits involved in each fermionic operation on an $32\times 32$ lattice when using an $n\times n$ Hybrid/Hybrid\+  mapping against $n$ in the context of a square-lattice qubit connectivity.The crosses denote the qubit-to-fermion ratio of the $n\times n$ Hybrid\+ mapping for each choice of $n$.}
    \label{fig:vary_cell_sizes_32}
\end{figure}
\begin{figure}[hbpt]
    \centering
    \includegraphics[width = 0.5\textwidth]{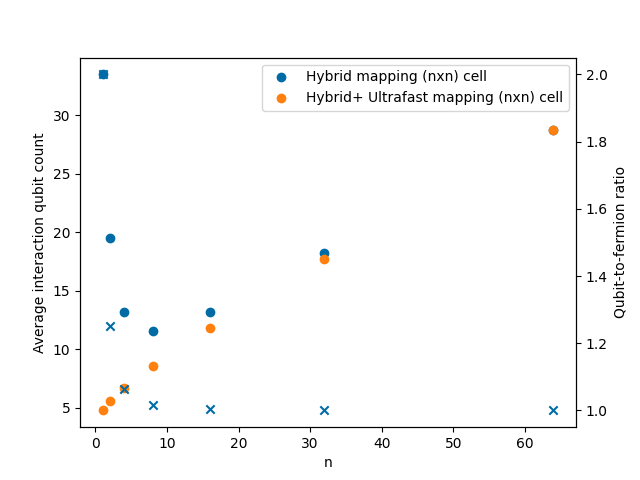}
    \caption{The average number of qubits involved in each fermionic operation on an $64\times 64$ lattice when using an $n\times n$ Hybrid/Hybrid\+  mapping against $n$ in the context of a square-lattice qubit connectivity for $n=\{1,2,4,8,16,32,64\}$. These results include qubits solely acted on by SWAP gates. The crosses denote the qubit-to-fermion ratio of the $n\times n$ Hybrid\+ mapping for each choice of $n$.}
    \label{fig:vary_cell_sizes_64}
\end{figure}
\begin{figure}[hbpt]
    \centering
    \includegraphics[width = 0.5\textwidth]{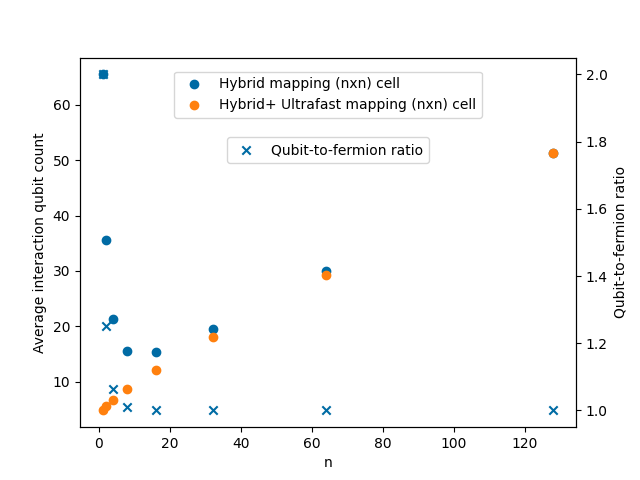}
    \caption{The average number of qubits involved in each fermionic operation on an $128\times 128$ lattice when using an $n\times n$ Hybrid/Hybrid\+  mapping against $n$ in the context of a square-lattice qubit connectivity for $n=\{1,2,4,8,16,32,64,128\}$. These results include qubits solely acted on by SWAP gates. The crosses denote the qubit-to-fermion ratio of the $n\times n$ Hybrid\+ mapping for each choice of $n$.}
    \label{fig:vary_cell_sizes_128}
\end{figure}
\subsection{Numerical Analysis}
\label{section:numerical}
We have considered the locality of each algorithm by computing the average interaction qubit count on a $N \times N$ square lattice. When investigating the case of a $4 \times 4$ cell, we found that the theoretical predictions held; the average interaction qubit count scaled with $\frac{N}{32}$ for the Hybrid algorithm and asymptotically approached a constant for the Hybrid\+ algorithm (as illustrated in Figure~\ref{fig:compare_mappings_ex_SWAP}).

We performed the same analysis in the context of an underlying square lattice connectivity. This required calculating the minimum number of SWAP gates required to connect all the qubits involved in the operation. We approximated the solution to this Steiner Tree problem using the SCIP-JACK library outlined in Section~\ref{section:SCIPJACK}. Figure~\ref{fig:compare_mappings} shows that the theoretical predictions hold with the average interaction qubit count increasing with $\frac{N}{8}$ for the Hybrid algorithm and tending to a constant for the Hybrid\+ algorithm.

The largest difference evident is in the Bravyi-Kitaev mapping. When including SWAP gates, the average interaction qubit count involved appears to be scaling linearly rather than logarithmically. In our calculations we used a trivial Z-pattern enumeration scheme for the Bravyi-Kitaev mapping, so it is possible the performance could be improved by an optimal choice of enumeration scheme.

Therefore in the context of an underlying square lattice connectivity the $4 \times 4$ Hybrid mapping out-performs both the Jordan-Wigner and Bravyi-Kitaev mapping by a factor of $4$ . Furthermore, even with all-to-all connectivity it still outperforms the Jordan-Wigner mapping by a factor of $16$, and beats the Bravyi-Kitaev mapping for lattices smaller than $164 \times 164$.

We have investigated which cell sizes give optimal performance. The behaviour shown in Figure~\ref{fig:vary_cell_sizes_ex_SWAP} and ~\ref{fig:vary_cell_sizes_32} matches with our theoretical predictions. For the Hybrid mapping, in the regime of $n \ll N$ the average interaction qubit count scales with $O(\frac{1}{2n^2})$ (for fixed $N$). However, this is quickly dominated by the $O(\log n)$ growth of the Bravyi-Kitaev mapping within each cell. Therefore, as the logarithm dominates for $K\frac{N}{2n^2} < \log n$ (for constant $K$), we hypothesis that the optimal choice of $n$ will scale with roughly $O(\sqrt{N})$. The trend of increasing optimal choice of $n$ with lattice size is demonstrated by the increasing optimal cell size from $4\times 4$ to $8\times 8$ to $16 \times 16$ when the lattice size is increased from $32 \times 32$ (Figure~\ref{fig:vary_cell_sizes_32}) to $64 \times 64$ (Figure~\ref{fig:vary_cell_sizes_64}) to $128 \times 128$ (Figure~\ref{fig:vary_cell_sizes_128}).

For the Hybrid\+ mapping, there is no $O(\frac{1}{2n^2})$ regime as there are no non-local gates acting on the roots. Therefore, the behaviour is dominated by the Bravyi-Kitaev mapping within the cell. This is reflected in the $O(\log n)$ growth of average interaction qubit count. However, the optimal cell size is not simply the smallest (equivalent to AQM \cite{steudtner2019quantum}). The main benefit of this mapping is in offering a direct trade-off between number of ancillas and interaction qubit count. As the current minimum qubit-to-fermion ratio achieved by a local mapping is 1.25 \cite{chen2022equivalence}, we have analysed the next possible cell size of $4 \times 4$ which achieves a qubit-to-fermion ratio of 1.0625. However, using larger cell sizes requires far fewer ancilla qubits. For example, an $8 \times 8$ Hybrid\+ mapping requires 94\% fewer ancillas than the super-compact mapping scheme \cite{chen2022equivalence}.

\section{Discussion}
We have presented two novel mappings: the Hybrid mapping achieves the shortest circuits of all known mappings without ancillas. The margin of improvement is significant. As demonstrated in our simulations of a real-world qubit architecture, it generates $75\%$ fewer interaction qubits than the Jordan-Wigner and Bravyi-Kitaev mappings. Additionally, as the lattice sizes increases the optimum cell size grows and the performance of the Hybrid mapping further improves.

For calculations on current or near-term quantum computers the Hybrid mapping demonstrates an improvement over the Bravyi-Kitaev and Jordan-Wigner mappings even with all-to-all connectivity. This holds for lattices up to $164 \times 164$ (Figure~\ref{fig:compare_mappings_ex_SWAP}). Lattices larger than this would require more than 27000 qubits to simulate, which is much larger than the current largest quantum computer (433 qubits~\cite{gambetta2020ibm}). Therefore, the Hybrid mapping will likely outperform the Bravyi-Kitaev and Jordan-Wigner mappings on real devices for the foreseeable future.

The Hybrid\+ mapping demonstrates a remarkable improvement on all existing local mapping schemes. It achieves similar levels of locality to the Derby-Klassen \cite{derby2021compact}, Super-compact \cite{chen2022equivalence}, and  AQM \cite{steudtner2019quantum} type mappings but expends only a small fraction of the ancillas in comparison. Our example of a $4 \times 4$ Hybrid\+ mapping required 75\% less ancillas than the local (super-compact) mapping with the lowest known ancilla requirements.

The Hybrid\+ mapping also provides a direct and simple trade-off between locality and ancilla qubits. Increasing the cell size $n$ increases the interaction qubit count linearly whilst the number of ancillas required decreases with $O(\frac{1}{n^2})$.

The approach we have taken in constructing both of these mappings need not be restricted to square-lattice qubit architectures. It should be possible to reproduce similar mappings for any architecture and corresponding fermionic interaction graph that admit a cellular decompisition.

We have considered a fermionic interaction graph which coincides with our architecture. It would be interesting to explore mapping an arbitrary fermionic interaction graph onto a given architecture using a similar Hybrid construct of connected Bravyi-Kitaev cells. In particular, we would like to understand how this mapping would translate into 3D, and whether this would require a 3D qubit connectivity or if cubic cells can be mapped to a planar grid efficiently.

In our analysis, we found connectivity constraints significantly hampered the Bravyi-Kitaev type terms in the mapping. Further work could explore this phenomenon in more detail. It is possible that an optimal choice of fermionic enumerations following the optimization techniques of~\cite{chiew2021optimal} could significantly reduce the impact of the connectivity constraints.

A lesser-known but equally remarkable role of fermion-to-qubit mappings is their role as duality transformations. The Jordan-Wigner transform~\cite{jordan1993paulische} can be regarded as an example of such a duality transform~\cite{henderson2022strong}. It maps spin Hamiltonians (such as XXZ chain) to simple fermionic systems which can be investigated by classical quantum chemistry techniques or by quantum means~\cite{nishimori2010elements,cao2019quantum}. Our two new families of fermion-to-qubit mappings could be interpreted as efficient duality transformations. Therefore, an interesting application of these hybrid mappings is the investigation of complex Hamiltonians.

{\it Acknowledgements} SS acknowledges support from the Royal Society University Research Fellowship and ``Quantum simulation algorithms for quantum chromodynamics'' grant (ST/W006251/1).

\bibliography{biblio}

\clearpage
\appendix
\onecolumngrid
\section{Visual guide to Hybrid mappings}
In this section we provide a series of diagrams to further elaborate and explain our hybrid mapping algorithms.\\\\
Figures~\ref{ZPattern} and \ref{SPattern} respectively describe the Jordan-Wigner mapping over the roots of Bravyi-Kitaev cells in the Hybrid and Hybrid\+ mapping schemes.\\\\
Figure~\ref{fig:stabilisers} illustrates the family of stabilisers utilised by the Hybrid\+ mapping scheme. This is identical to the stabilisers found in \cite{steudtner2019quantum}.\\\\
Figures~\ref{fig:verticalInteractionYodd} and \ref{fig:horizontalInteractionXYYX} demonstrate how the Hybrid\+ mapping scheme shortens different Pauli-strings.\\\\
In the figures below: the dashed boxes represent the cells, whilst the circles inside them represent the roots of each cell, and the circles without boxes represent the ancillas.
\renewcommand\thefigure{\thesection.\arabic{figure}}
\setcounter{figure}{0}    
\label{appendix}
    \begin{figure}[h]
    \begin{minipage}{0.5\textwidth}
        \centering
         \includegraphics[page=20, width=\textwidth  , height=5.75cm]{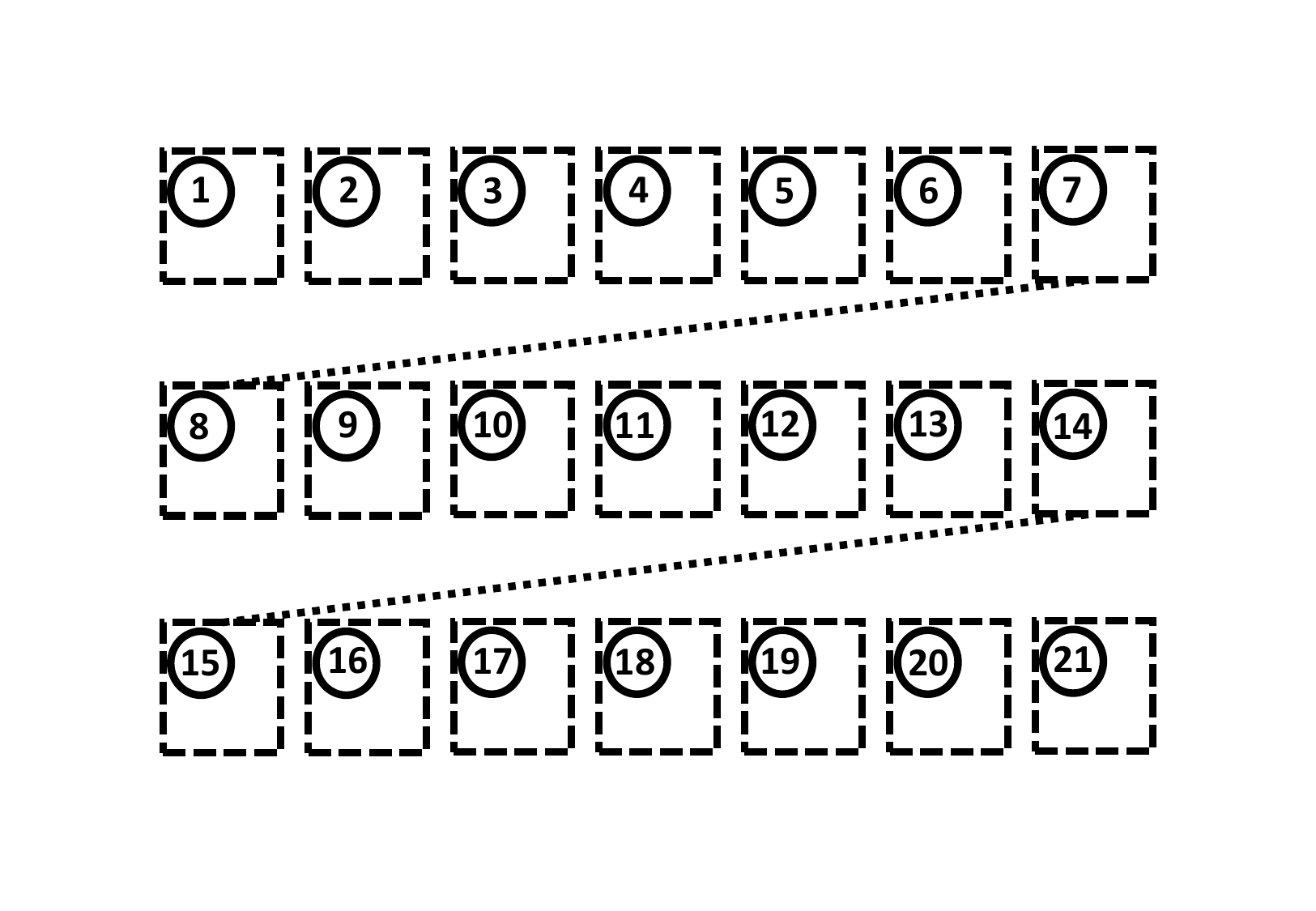}
        \caption{ \label{ZPattern}Diagram illustrating how the \textit{cell number} is assigned in a Z pattern to each cell in the Hybrid mapping.}
        \end{minipage}%
        \begin{minipage}{0.5\textwidth}
        \centering
         \includegraphics[page=19, width=\textwidth  , height=5.75cm]{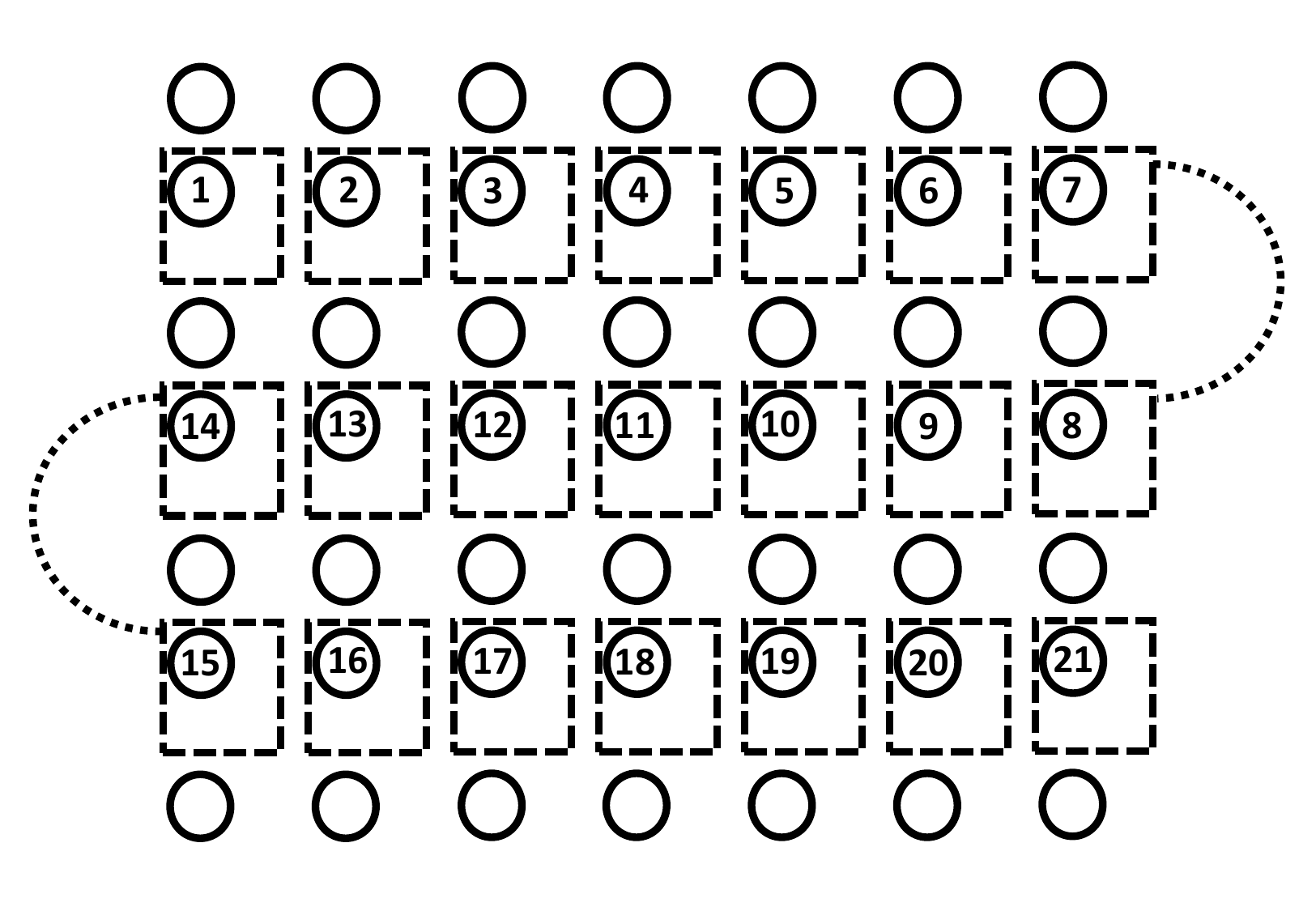}
        \caption{\label{SPattern}Diagram illustrating how the \textit{cell number} is assigned in a S pattern to each cell in the Hybrid\+ mapping. The unnumbered circles represent the ancillas each connected to the cell directly below them.}
        \end{minipage}
    \end{figure}
            \begin{figure*}[btp]
        \centering
        \begin{subfigure}{.5\textwidth}
        \includegraphics[page=2, width=\textwidth  , height=6cm]{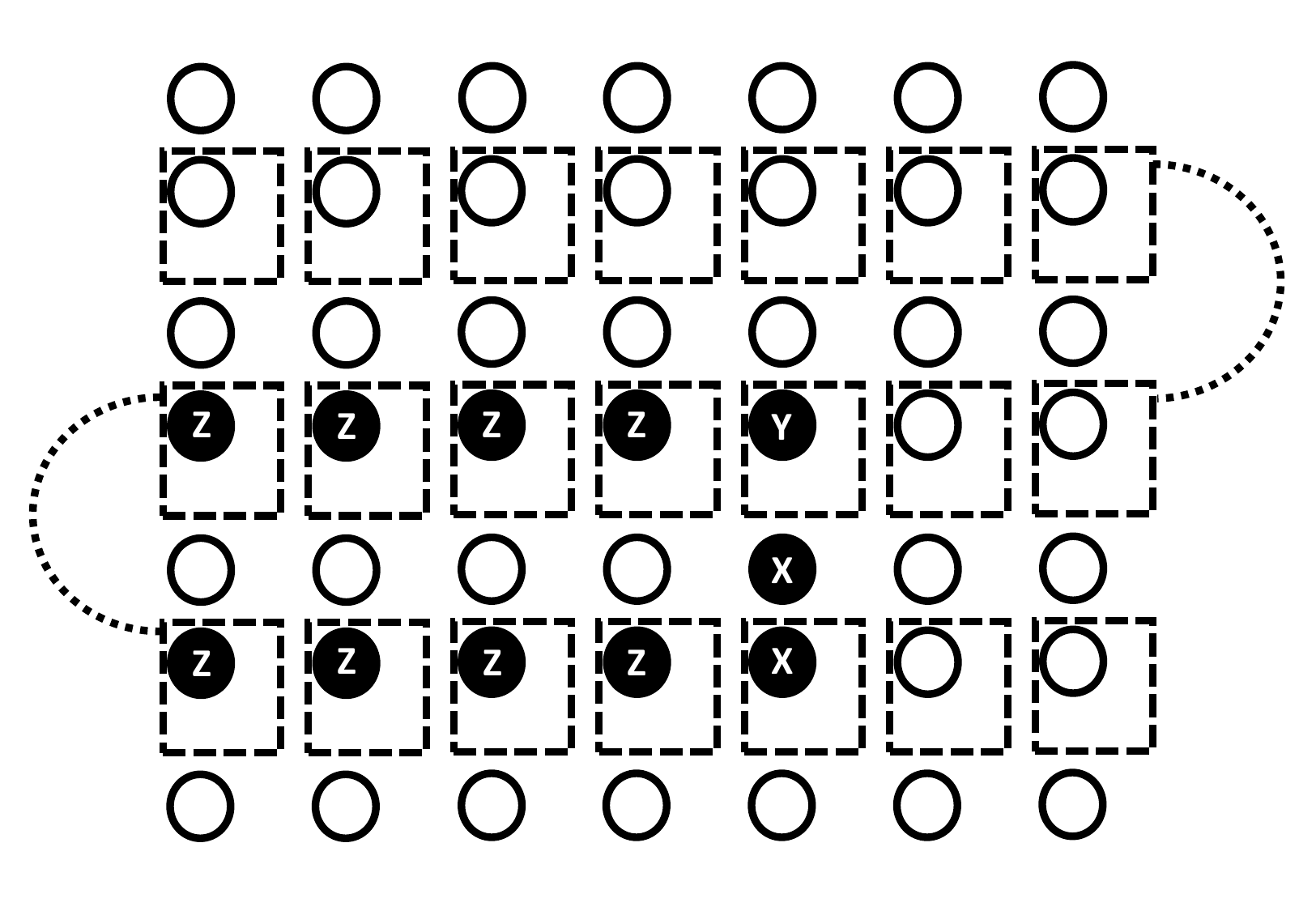}
        \end{subfigure}%
        \begin{subfigure}{.5\textwidth}
        \includegraphics[page=3, width=\textwidth  , height=6cm]{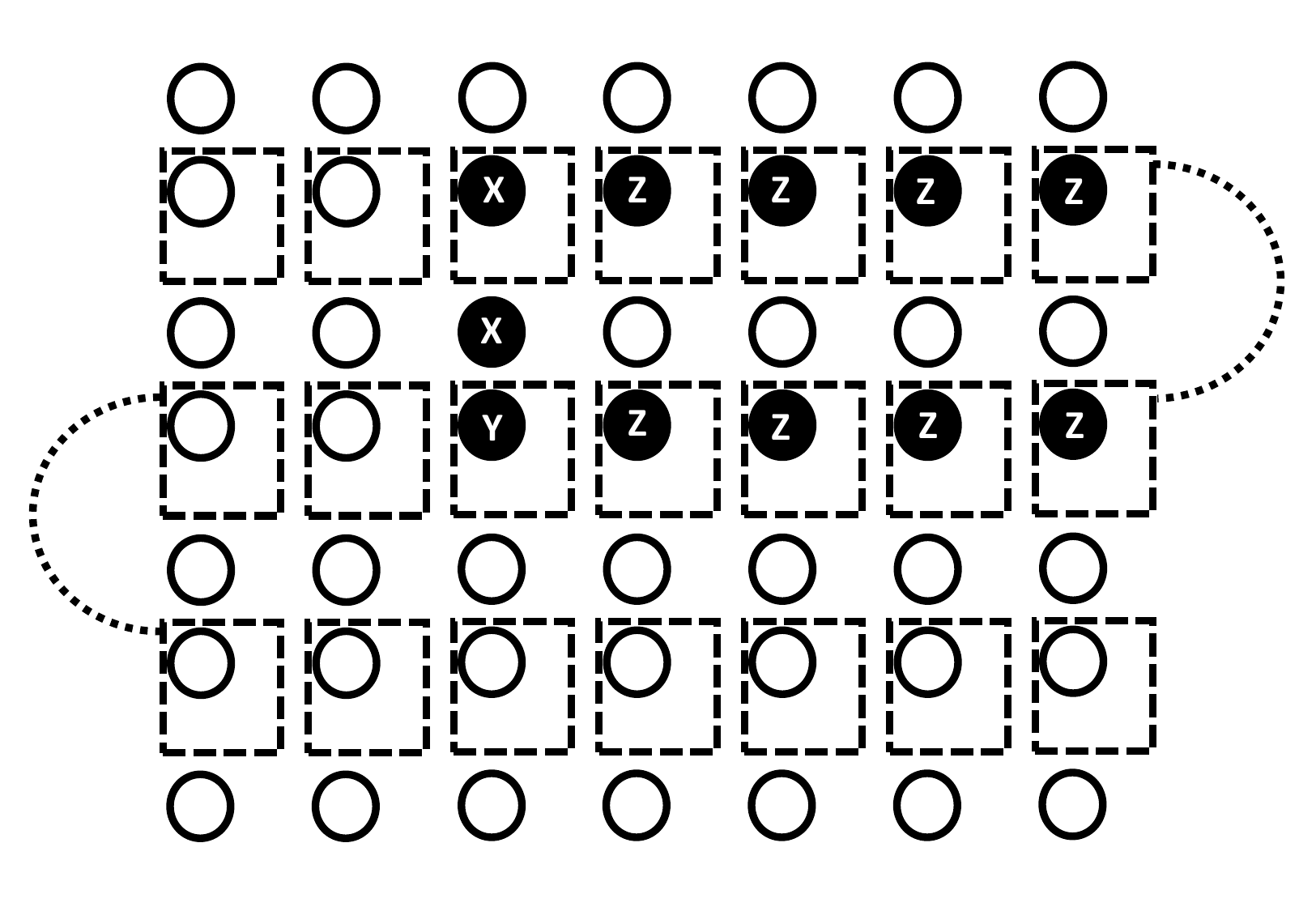}
        \end{subfigure}%
        
                \caption{\label{fig:stabilisers} Examples from the family of stabilisers created by applying the series of gates illustrated in \ref{fig:entangleEdge} and \ref{fig:entangleRow}. The family consists of a stabiliser for every ancilla which is a string of Z gates that follow the line of the S-pattern with either $XXY$ or $YXX$ gates surrounding the ancilla depending on whether the row is odd or even.}
    \end{figure*}
       \begin{figure*}
       \begin{minipage}{\textwidth}
           \begin{minipage}{0.5\textwidth}
        \includegraphics[page=9, width=\textwidth  , height=5.75cm]{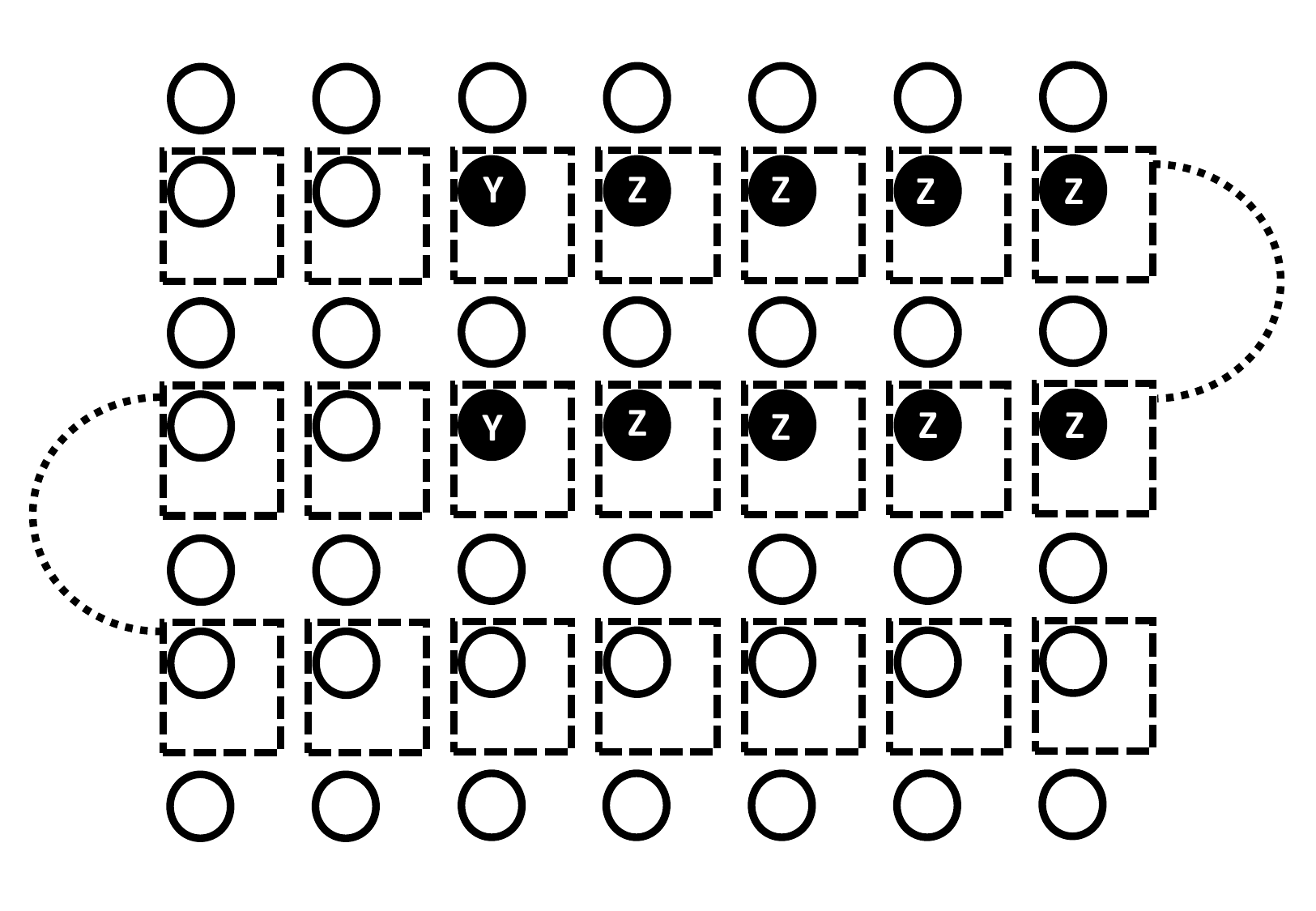}
        \tikz[baseline=-\baselineskip]\draw[ultra thick,->] (0,0) -- ++ (0,-0.5);
        \includegraphics[page=10, width=\textwidth  , height=5.75cm]{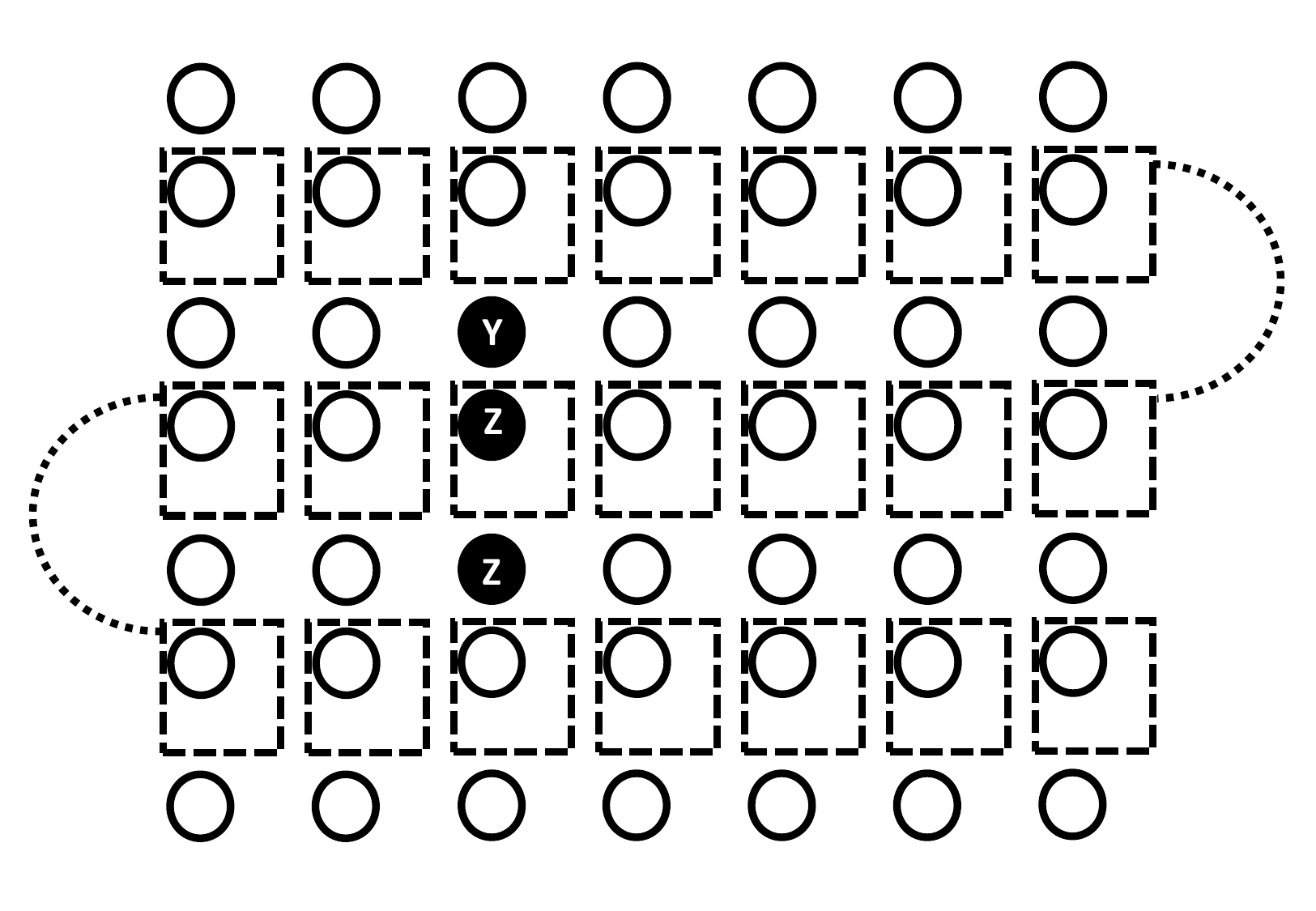}
        \end{minipage}%
    \begin{minipage}{0.5\textwidth}
        \includegraphics[page=11, width=\textwidth  , height=5.75cm]{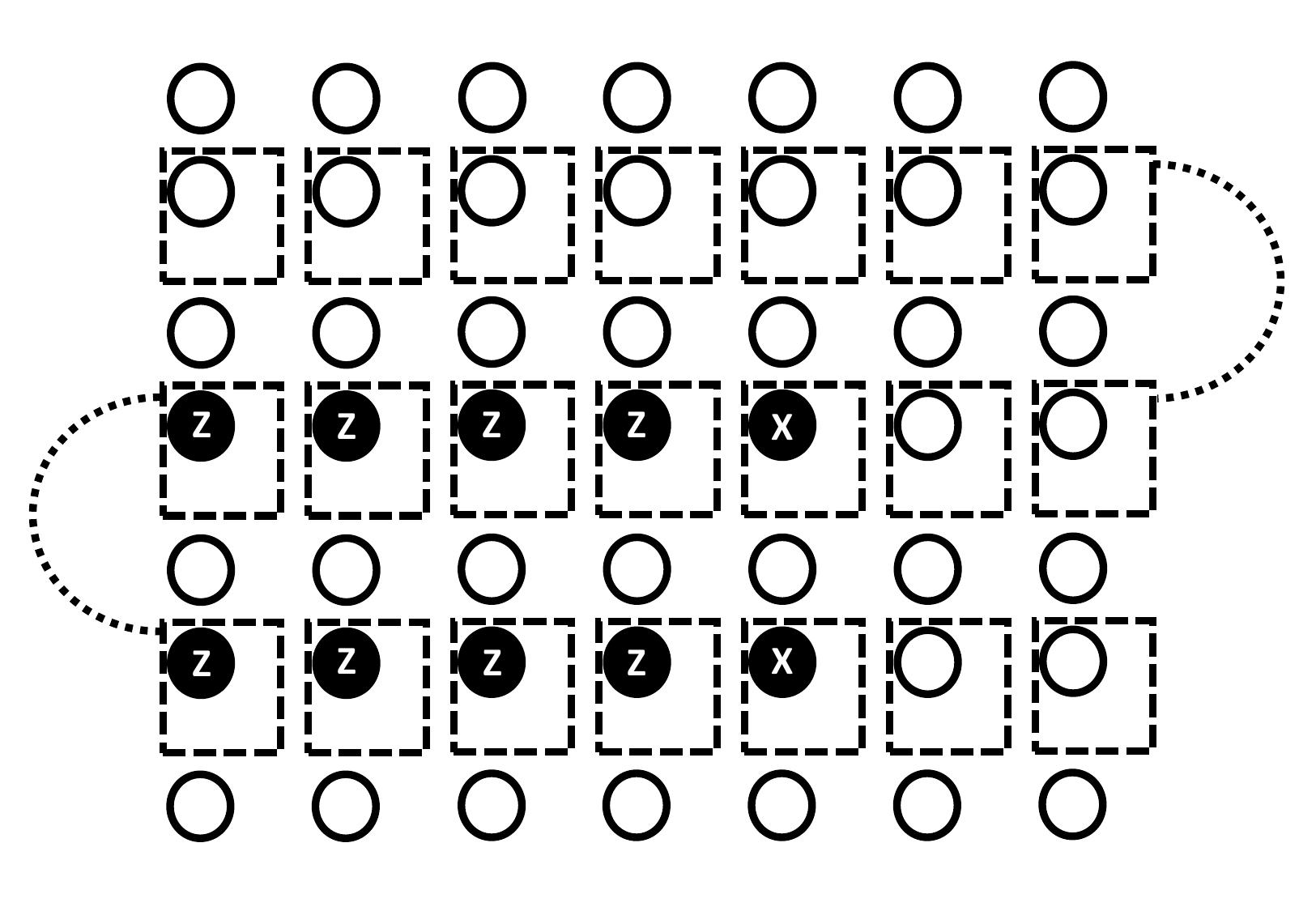}
        \tikz[baseline=-\baselineskip]\draw[ultra thick,->] (0,0) -- ++ (0,-0.5);
        \includegraphics[page=12, width=\textwidth  , height=5.75cm]{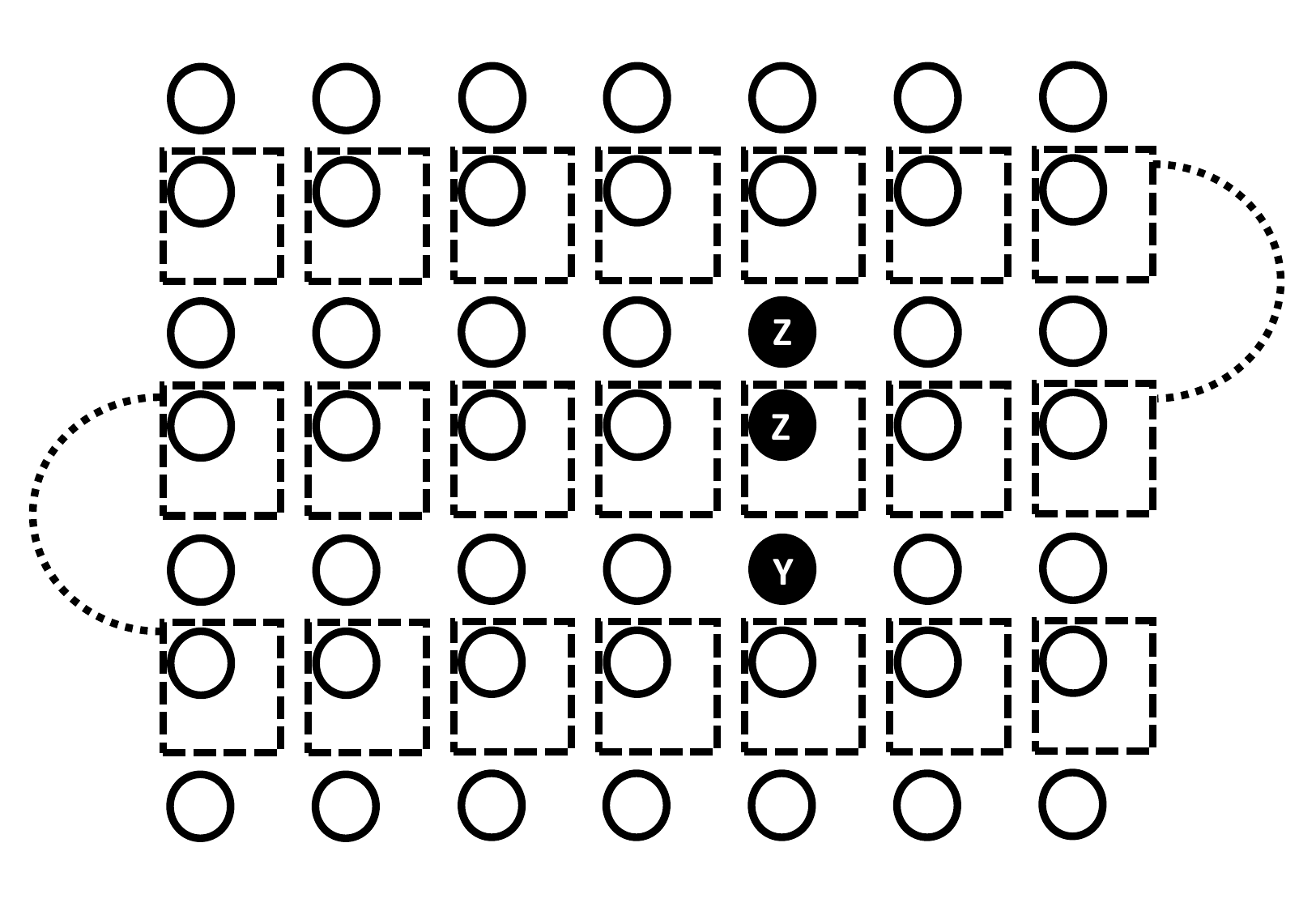}
        \end{minipage}
        \begin{minipage}{0.5\textwidth}
        \includegraphics[page=13, width=\textwidth  , height=5.75cm]{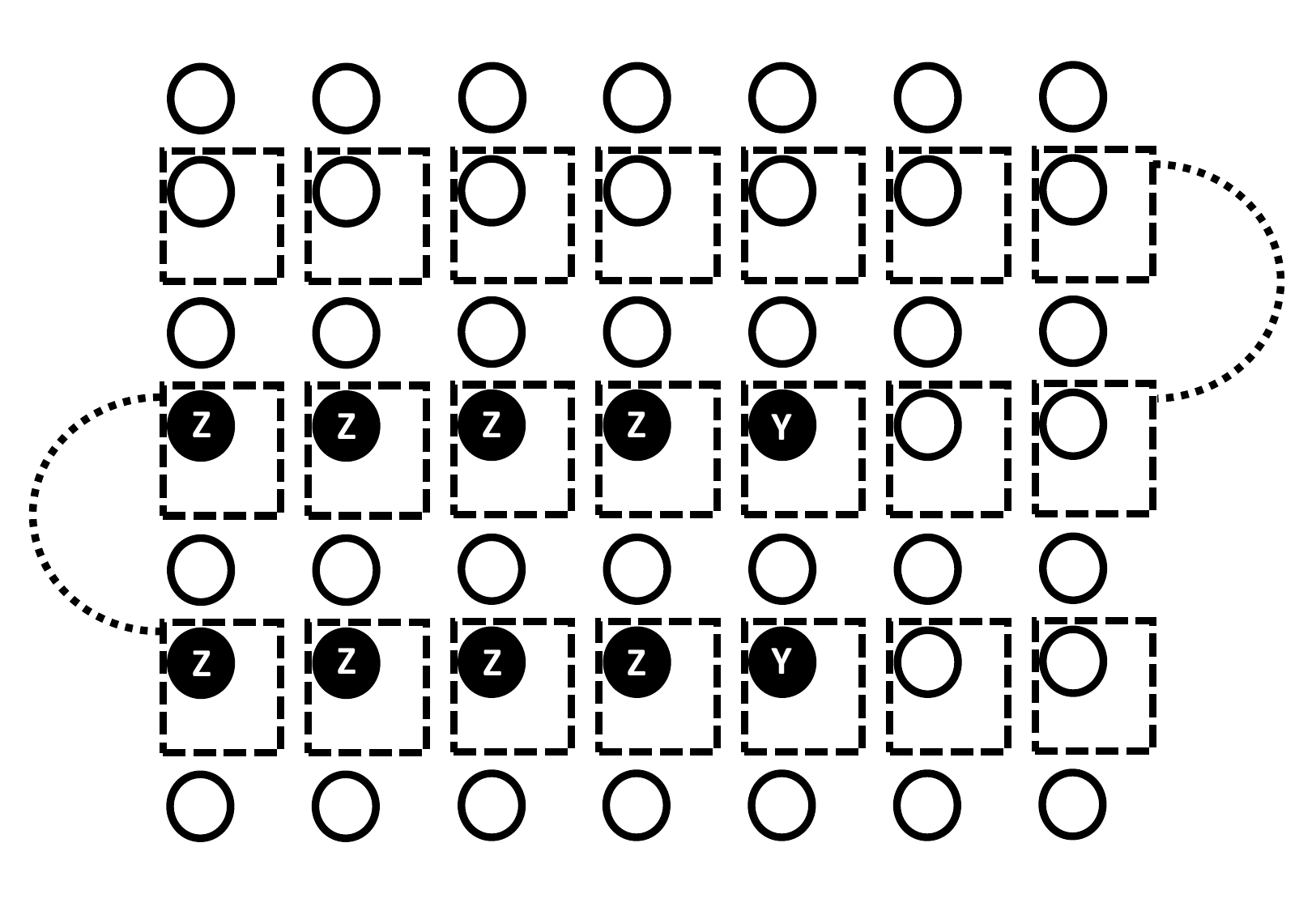}
        \tikz[baseline=-\baselineskip]\draw[ultra thick,->] (0,0) -- ++ (0,-0.5);
        \includegraphics[page=14, width=\textwidth  , height=5.75cm]{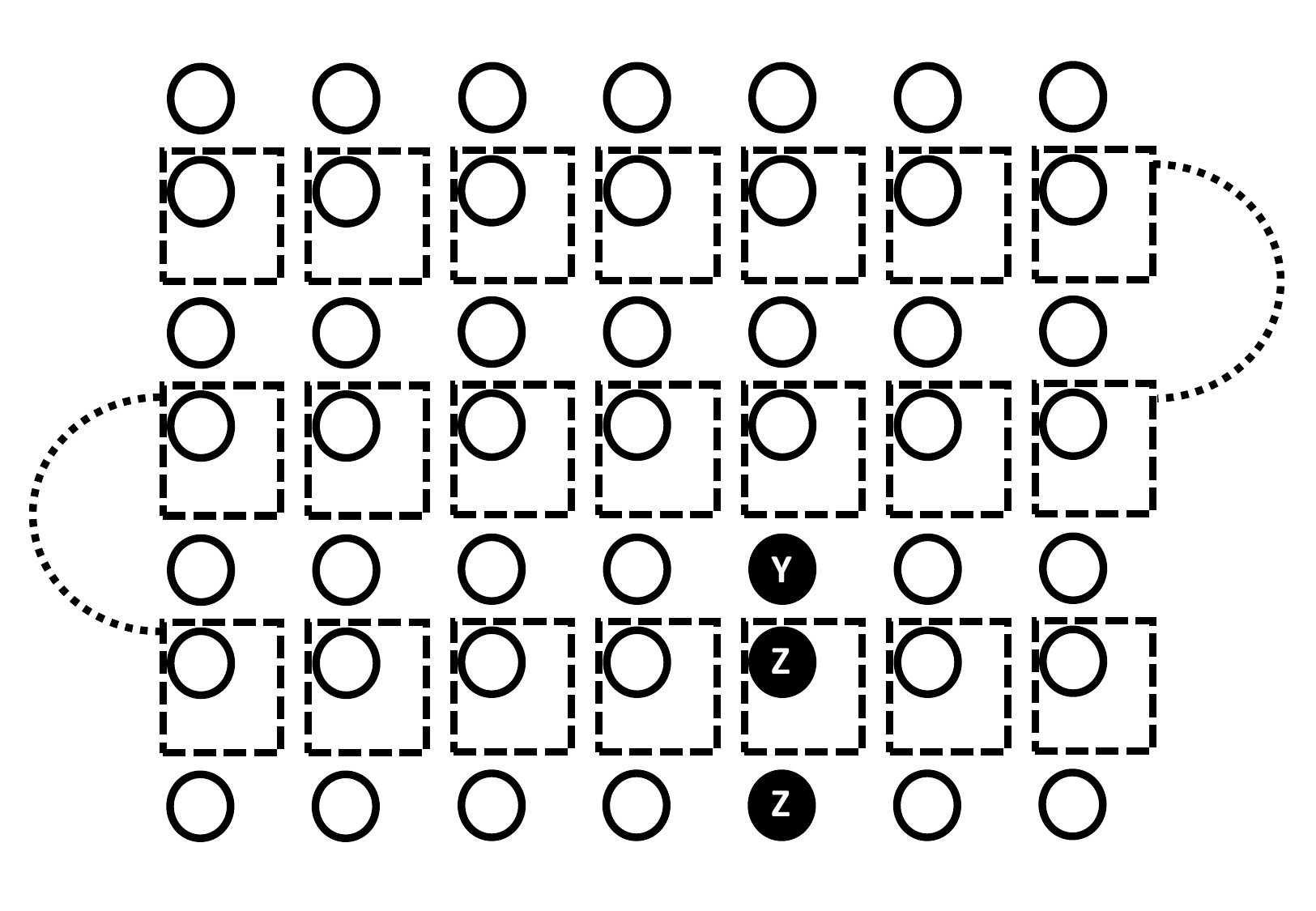}
\end{minipage}%
        \begin{minipage}{0.5\textwidth}
        \includegraphics[page=15, width=\textwidth  , height=5.75cm]{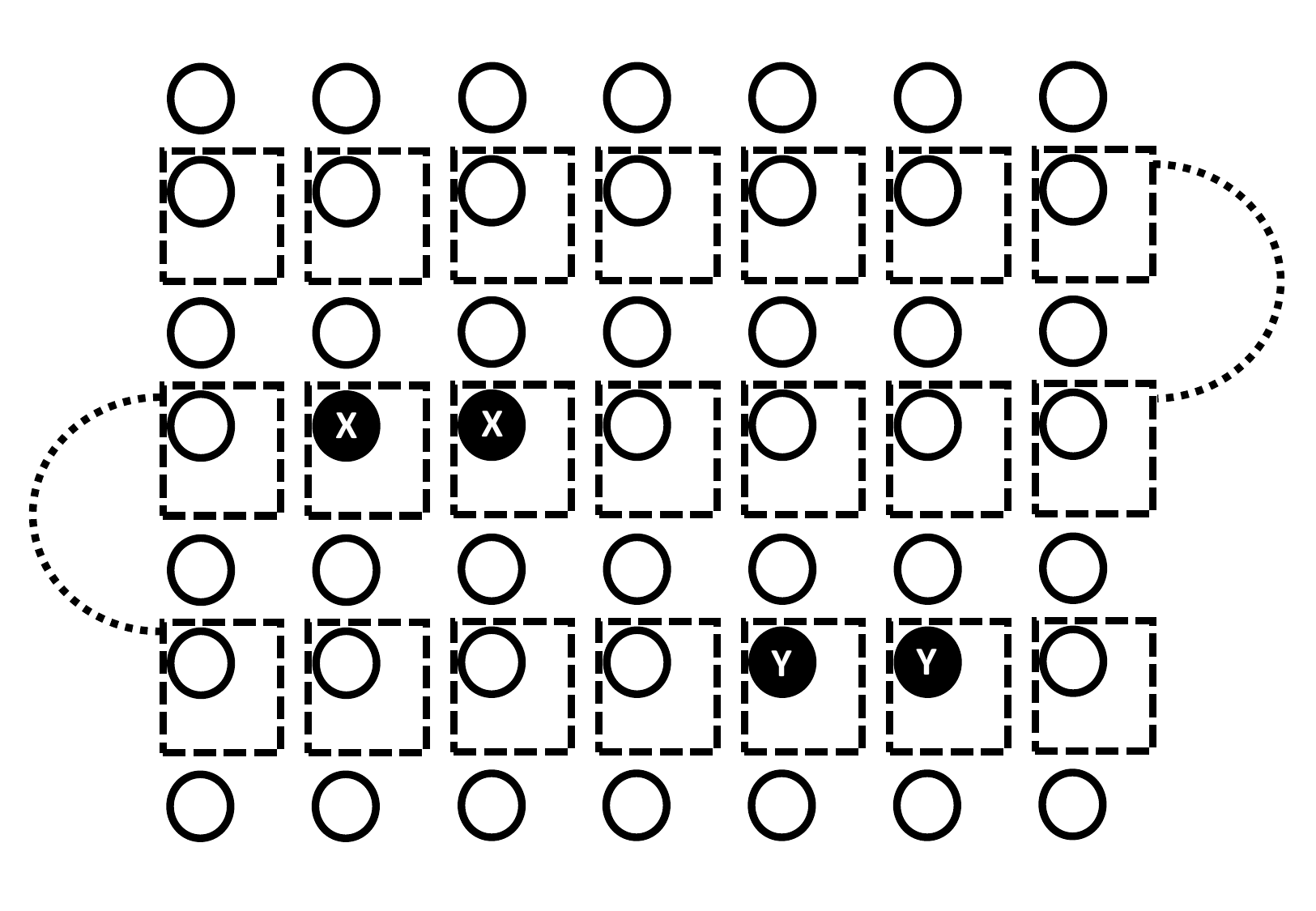}
        \tikz[baseline=-\baselineskip]\draw[ultra thick,->] (0,0) -- ++ (0,-0.5);
        \includegraphics[page=16, width=\textwidth , height=5.75cm]{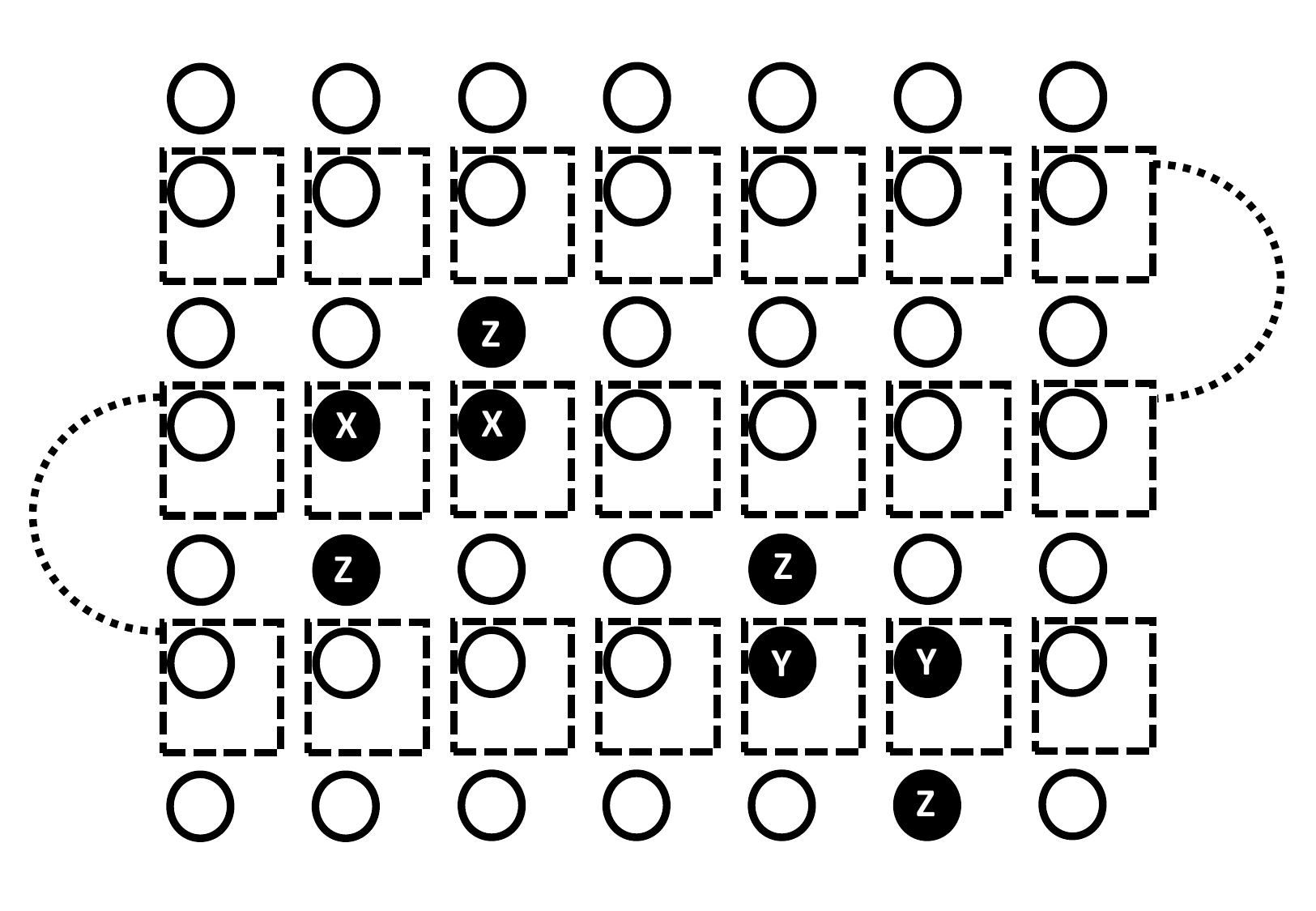}
\end{minipage}
\end{minipage}
        \caption{\label{fig:verticalInteractionYodd}Diagrams showing how the portion of Pauli-strings acting on the roots of cells are modified and simplified by the Hybrid\+ mapping algorithm}
    \end{figure*}
\begin{figure}
\centering
 \begin{minipage}{0.5\textwidth}
        \includegraphics[page=17, width=\textwidth  , height=6cm]{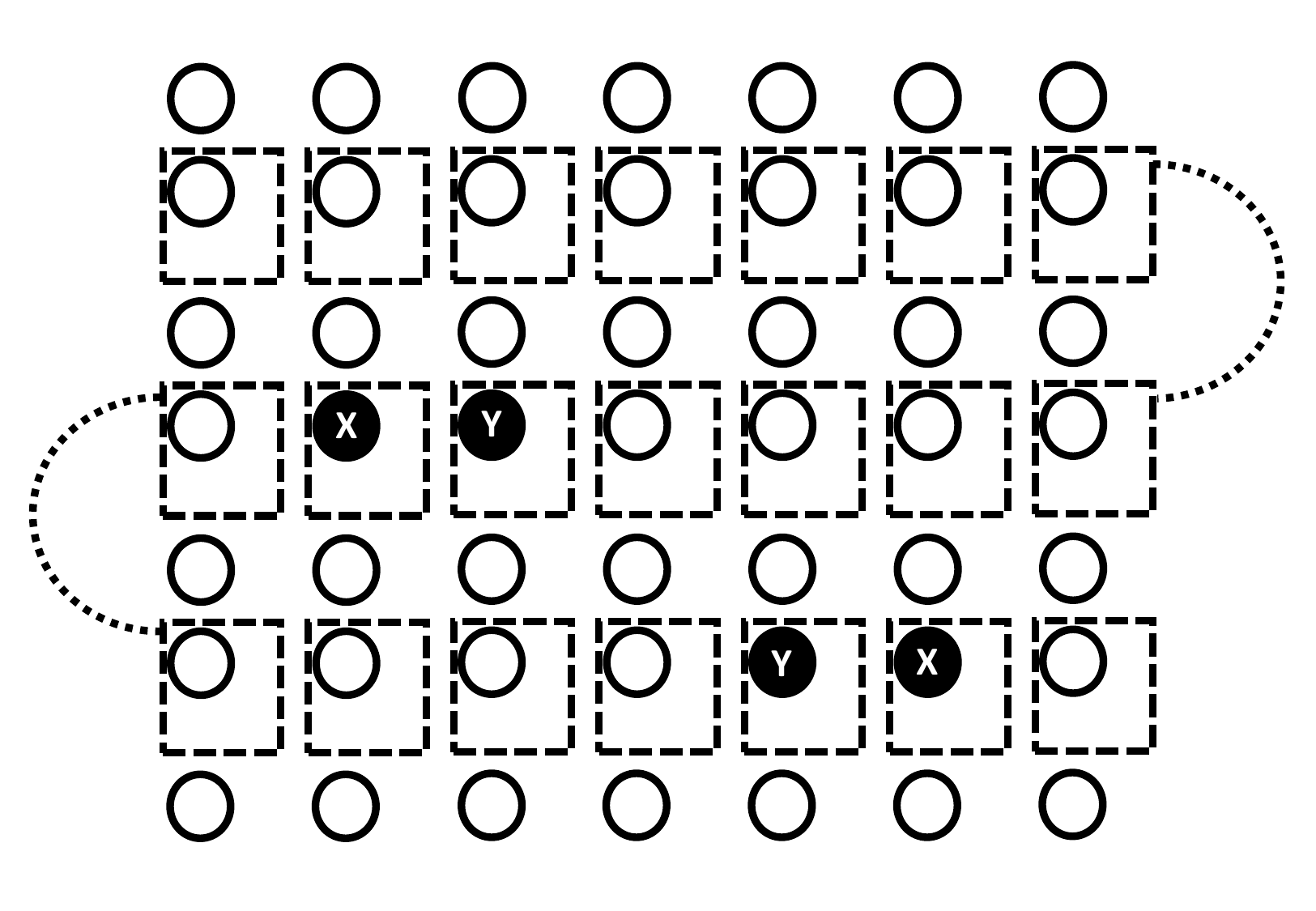}
        \tikz[baseline=-\baselineskip]\draw[ultra thick,->] (0,0) -- ++ (0,-0.5);
        \includegraphics[page=18, width=\textwidth  , height=6cm]{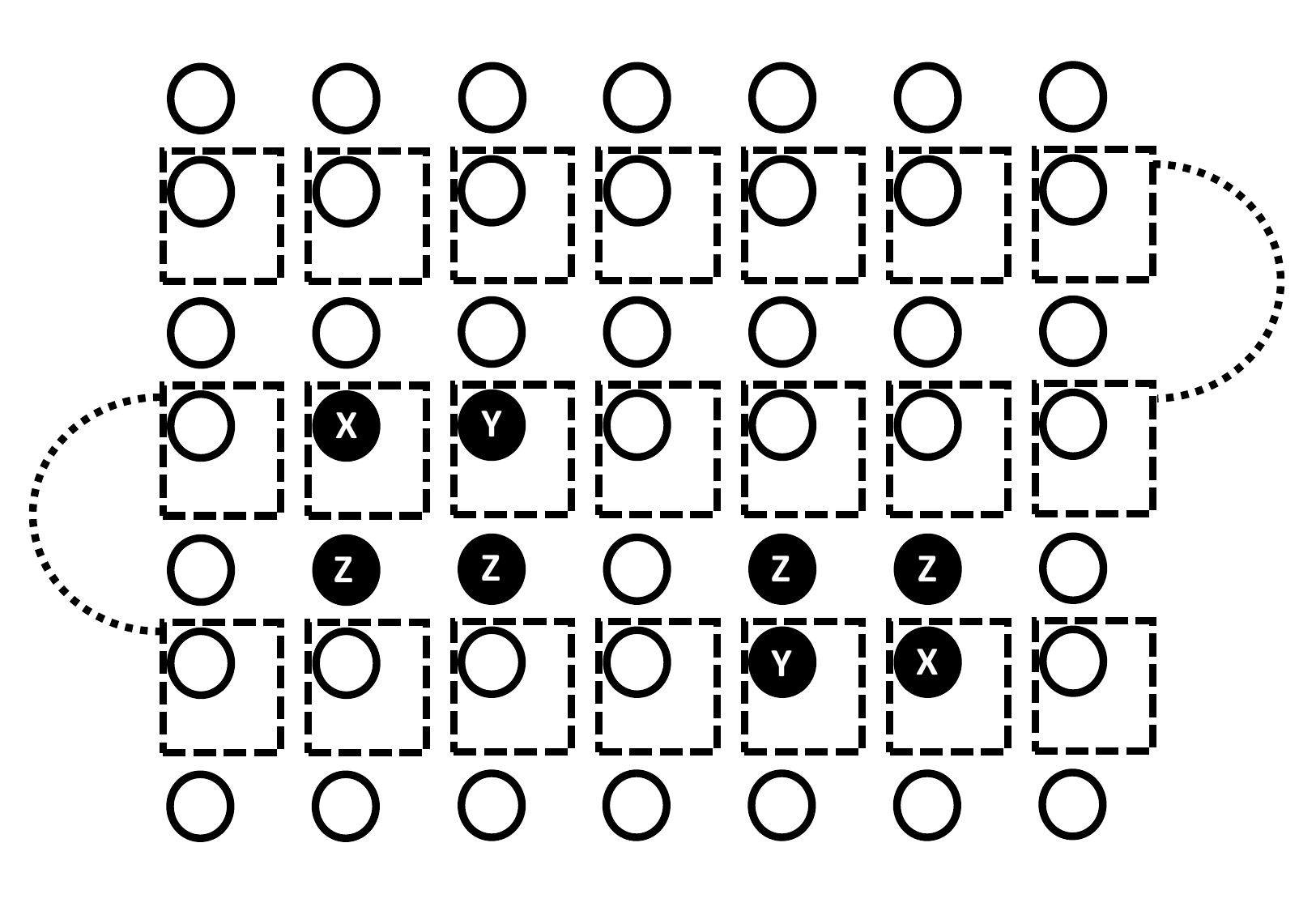}
        \end{minipage}
        \caption{\label{fig:horizontalInteractionXYYX}Diagrams showing how the portion of a Pauli-string acting on the roots of the cells is modified and simplified by the Hybrid\+ mapping algorithm}
    \end{figure}
\end{document}